\documentclass[aps,superscriptaddress,twocolumn,prd]{revtex4-2}

\usepackage[utf8]{inputenc}
\usepackage{amsmath,amssymb}
\usepackage{graphicx}
\usepackage{hyperref}
\usepackage{xcolor}
\usepackage{orcidlink}

\usepackage{axodraw2}
\usepackage{color}
\usepackage{amssymb}
\usepackage[normalem]{ulem}
\usepackage{graphicx,color}
\usepackage{amsmath}
\usepackage{subfigure}
\usepackage[normalem]{ulem}
\usepackage{booktabs}
\usepackage{xcolor}
\usepackage{soul}
\usepackage{epsfig,graphics,color,graphicx,amsmath}
 \usepackage{epstopdf}
  \usepackage{epsfig,graphics,color,graphicx,amsmath}

\usepackage{tikz}

\usepackage{mathtools}

\def\be{\begin{equation}}
\def\ee{\end{equation}}
\def\bea{\begin{eqnarray}}
\def\eea{\end{eqnarray}}
\def\psla{\slash \! \!\! }

\colorlet{darkgreen}{green!50!black}
\colorlet{brightyellow}{yellow!75!red}
\colorlet{orange}{red!50!yellow}
\colorlet{darkblue}{blue!60!black}
\colorlet{darkred}{red!80!black}
\usepackage{array}
\usepackage{array,etoolbox}
\usepackage{fancyhdr}
\usepackage{slashed}

\def\sla#1{\rlap\slash #1}

\newcommand{\bno}{\begin{eqnarray*}}
\newcommand{\eno}{\end{eqnarray*}}

\newcommand{\bl}{\begin{large}}
\newcommand{\el}{\end{large}}
\newcommand{\bla}{\begin{Large}}
\newcommand{\ela}{\end{Large}}

\newcommand{\ede}{{\end{document}}}

\def\be{\begin{equation}}
\def\ee{\end{equation}}
\def\bea{\begin{eqnarray}}
\def\eea{\end{eqnarray}}

\begin{document}


\title{Light-front mass operator with  dressed quarks }

\author{J. A. O. Marinho\orcidlink{0000-0003-1897-3276}}
\email{adnei@ufra.edu.br}
\affiliation{Universidade Federal Rural da Amazônia, Campus de Parauapebas, Av Duane Silva Sousa, 68.515-000}
\affiliation{ Parauapebas, PA, Brazil, Programa de P\'os-graduaç\~ao da Faculdade de F\'isica. Universidade Federal do Sul e Sudeste do Para. 68500-000, Marab\'a- PA, Brazil}

\author{J. P. B. C. de Melo\orcidlink{0000-0003-1551-9609}}
\email{
joao.mello@cruzeirodosul.edu.br}
\affiliation{Laborat\'orio de F\'\i sica Te\'orica e Computacional-LFTC, 
Universidade Cruzeiro do Sul / Universidade Cidade de S\~ao Paulo,
015060-000, S\~ao Paulo, SP, Brazil}

\author{T. Frederico\orcidlink{0000-0002-5497-5490}
} 
\email{tobias@ita.br}

\affiliation{Instituto Tecnol\'ogico de Aeron\'autica, 12.228-900, S\~ao Jos\'e dos Campos, SP, Brazil} 
 
\author{W. de Paula\orcidlink{0000-0002-7701-0421}}
\email{wayne@ita.br}
\affiliation{Instituto Tecnol\'ogico de Aeron\'autica, 12.228-900, S\~ao Jos\'e dos Campos, SP, Brazil}

\begin{abstract} 
We construct an effective light-front mass-squared operator for quark–antiquark systems that incorporates quark dressing effects through a running quark mass. Starting from a Minkowski-space quark propagator constrained by lattice-QCD–inspired parametrization, we derive the disconnected light-front resolvent for a light quark-antiquark system using a generalized spectral representation of an individual quark propagator separating out the instantaneous contributions. By projecting the resolvent onto a constituent-quark helicity basis, we obtain an effective dressed mass-squared operator suitable for light-front Hamiltonian approaches. We introduce an effective light-front quark self-energy and analyze its momentum dependence. As an application, we study pion  structure using representative light-front wave-function models and compute unpolarized transverse-momentum–dependent distributions, unpolarized parton distribution functions and distribution amplitudes. Our results show that quark dressing induces sizable infrared modifications while preserving controlled ultraviolet behavior, providing a framework to incorporate nonperturbative QCD dynamics into light-front descriptions of hadrons.
\end{abstract}
\date{\today}
\maketitle

\section{Introduction}
The strong interactions presents the persistent challenge of describing Quantum Chromodynamics (QCD) consistently across its infrared (IR) and ultraviolet (UV) regimes. In particular, the nonperturbative sector of QCD remains one of the most demanding frontiers in theoretical physics. A comprehensive understanding of hadronic structure and dynamics requires nonperturbative tools, such as functional methods relevant in the IR domain, together with perturbative techniques based on integral equations that describe the perturbative expansion of the theory \cite{Eichmann:2016yit}. Consistency between the results obtained from functional approaches and those derived from Schwinger–Dyson and Bethe–Salpeter equations, formulated in both Euclidean and Minkowski spaces, is therefore essential \cite{dePaula:2026gtb}. When combined with lattice QCD simulations, these continuum approaches allow for a unified and consistent treatment of QCD Green’s functions and bound-state dynamics.

Significant progress has been achieved in the study of the quark–gluon vertex \cite{Oliveira:2018fkj, Oliveira:2018ukh, Oliveira:2020yac, Aguilar:2024ciu} and its impact on the Bethe–Salpeter kernels for mesons and baryons \cite{Xu:2022kng}. In particular, Minkowski space formulations have led to recent advances in understanding the dressed quark~\cite{Sauli:2006ba,Mezrag:2020iuo,Duarte:2022yur}, the pion~\cite{dePaula:2020qna,Ydrefors:2021dwa,dePaula:2022pcb,dePaula:2023ver}, and the nucleon~\cite{Ydrefors:2022bhq}. These developments provide the necessary groundwork for constructing effective descriptions of hadrons directly in Minkowski space.

Lattice QCD simulations~\cite{Oliveira2019,Li:2019hyv} and continuum studies based on Schwinger–Dyson equations~\cite{Aguilar:2022wsh} have demonstrated that QCD dynamically generates momentum-dependent mass scales for quarks and gluons. These scales dominate the infrared behavior of the theory and are closely related to confinement~\cite{Dudal:2010tf,Cucchieri:2011ig}. They also account for the dominant contribution to the nucleon mass~\cite{Roberts:2021nhw}, and hence to the mass of visible matter in the Universe, beyond the Higgs mechanism. Understanding how these dynamically generated masses are incorporated into effective descriptions of hadrons is a central objective of current and future experimental programs, such as the Electron–Ion Collider (EIC)~\cite{Accardi:2023chb}.

The emergence of hadron masses is associated with nontrivial configurations involving infinitely many dressed quarks and gluons, which are described by the Schwinger–Dyson, Bethe–Salpeter, and Faddeev equations~\cite{Eichmann:2016yit}. This description relies on the infrared behavior of QCD Green’s functions, where dynamical chiral symmetry breaking (DCSB) generates constituent-like quark masses and leads to the appearance of the pion and kaon as (pseudo-)Goldstone bosons~\cite{Arrington_2021}. In this regime, dressed quarks and gluons emerge as the relevant effective degrees of freedom underlying both hadron spectroscopy and structure calculations, including those performed within the basis light-front quantization (BLFQ) framework~\cite{Lan:2019vui}.

In Minkowski space, hadrons reflect the full complexity of QCD dynamics. The light-front (LF) wave function contains an infinite tower of Fock components dynamically coupled by the interaction~\cite{bakker2014light}. However, Haag’s theorem~\cite{Haag:1955ev,Streater:1989vi} implies that interacting quantum field theories cannot be constructed in the same Hilbert space as free theories, posing conceptual challenges for Hamiltonian formulations of QCD. A possible resolution was proposed in Ref.~\cite{Polyzou:2021qpr}, where a dual Hilbert-space framework is introduced, allowing effective one-particle states to be defined from the interacting theory itself.

Although quarks and gluons are confined and do not appear as asymptotic states, their propagators can be investigated through generalized Källén–Lehmann representations~\cite{itzykson2012quantum}, constrained by lattice QCD data in the spacelike region and analytically continued to the timelike domain~\cite{Li:2019hyv}. Combined with light-front projection techniques~\cite{Frederico:2010zh}, these representations allow the construction of effective light-front resolvents for dressed quarks \cite{dePaula:2026gtb}. These resolvents naturally enter in the diagonal part of the light-front mass operator in the Fock space and provide a systematic way to incorporate quark dressing effects into light-front Hamiltonian formulations, such as BLFQ\cite{Vary:2009gt,Vary:2016ccz}, and into light-front projections of the Bethe–Salpeter equation~\cite{dePaula:2026gtb}.

The goal of the present work is to develop an effective light-front formulation in which quark dressing effects, encoded in a momentum-dependent quark mass, are consistently incorporated into the squared mass operator of light quark–antiquark systems. Starting from a phenomenological quark propagator model constrained by lattice QCD results, we construct the disconnected light-front resolvent using a generalized spectral representation of the quark propagator  separating out the instantaneous term. By projecting this resolvent onto a quark helicity basis with a constituent mass defined by the infrared value of the running quark mass, we derive an effective dressed mass-squared operator in place of the usual free mass-squared operator. This effective dressed mass-squared operator is expected to be suitable for practical applications of the light-front Hamiltonian for light hadrons systems. Within this framework, we further define an effective light-front quark self-energy that captures dressing effects and ensures the correct ultraviolet behavior dictated by asymptotic freedom.

As a concrete application, we analyze the pion structure within this light-front dressed quark framework. Using representative Gaussian and power-law light-front wave-function models, we investigate how the running quark mass affects unpolarized transverse-momentum–dependent distributions, unpolarized parton distribution functions, and distribution amplitudes. This allows to assess the phenomenological impact of the quark dressing and dynamical chiral symmetry breaking effects on pion observables, while using models parametrized with  the pion decay constant and valence probability.

This work is organized as follows. In Sec.~\ref{Sec:DiracStr} we discuss the Dirac structure of the dressed quark propagator and the implementation of running quark masses. Section~\ref{sec:QuarkProp} introduces the quark propagator model employed in this work. In Sec.~\ref{sec:DressedMO} we construct the dressed light-front mass operator, while Sec.~\ref{sec:ELCQSE} is devoted to the derivation of the effective light-front quark self-energy. Pion phenomenology is presented in Sec.~\ref{sec:PionPheno}. Finally, Sec.~\ref{sec:Summa} summarizes our results and outlines perspectives for future developments.

\section{The Dirac Structure and The Running Masses} \label{Sec:DiracStr}

A dressed Dirac propagator, for a parity-conserving interaction, can be written in a general form as: 
\begin{multline}
    S(k)= S^V(k^2) \psla k + S^S(k^2)=
\\ =
\imath \int_{0}^{\infty}ds~\frac{\rho^V(s)\, \psla k+\rho^S(s)}{k^2-s+ \imath\varepsilon }~,
    \label{eq:KLR}
\end{multline}
where in the first line one has the decomposition in terms of allowed Dirac structures and  corresponding scalar functions,  while in the second line one has the dispersive representation (see, e.g., Ref.~\cite{itzykson2012quantum} for the K\"allen-Lehman (KL) representation in  QED).  For QCD it is not yet established from ab-initio that such an integral representation is valid (see the discussion on e.g.~\cite{Horak:2022aza}).

The decomposition of this propagator naturally separates the so-called  ``instantaneous terms'' in the light-front framework. These terms emerge from the decomposition of the fermion propagator in terms of light-front variables, notably due to the presence of the constrained field component $( \psi_- )$, which does not propagate in light-front time. 
The first explicit derivation and interpretation of such contributions can be found in the seminal work of Yan \cite{Yan:1973}, who showed that the light-front fermion propagator contains, besides a propagating part in the LF time, an additional term proportional to $\gamma^+/2k^+$, referred to as the instantaneous term~\cite{deMelo1998an,Bhamre2025}. This term can give rise to contributions to structure observables that are non-diagonal in the light-front Fock space, such as electromagnetic form factors (see, e.g., \cite{Naus1997,deMelo1997,deMelo1997rho,
 deMelo2025a,deMelo2025a}). Our metric is such that $v^{\pm} = v^0 {\pm} v^3$, and $\mathbf{v}_\perp = \{v^1, v^2 \}$.

Such instantaneous contributions have been systematically explored in the context of relativistic bound-state equations, including the light-front projection of the Bethe-Salpeter equation \cite{dePaula:2016oct,dePaula:2017ikc}. Notably, Frederico and collaborators have developed a comprehensive framework for treating these terms using projected Green’s functions and quasi-potential equations \cite{Sales:2001prc}. Marinho et al. \cite{Marinho:2008pe} extended this formalism to cop with the constraints of electromagnetic current conservation.

We now separate the instantaneous terms that emerge in the propagators given in Eq. (\ref{eq:KLR}). Let us analyze the propagator:
\begin{multline}
    S(k)= 
\imath\int_{0}^{\infty}ds~\frac{\rho^V(s)\, \psla k_{(s)}^{\text{on}}+\rho^S(s)} {k^2-s + \imath\varepsilon } \\
+\imath\frac{\gamma^+}{2\,k^+}\int_{0}^{\infty}ds~\rho^V(s)
~,
    \label{eq:KLR1}
\end{multline}
where the LF components of the on-mass shell momentum,  $k^{on}_{(s)}$, are given by:
\begin{equation}
k_{(s)}^{\text{on},-}=\frac{\mathbf{k}_\perp^2+s}{k^+}~.
\label{eq:kon}
\end{equation}
where $s$ represents any squared mass.

The dressed quark two-point function in  propagating LF-time is the propagator from Eq.(\ref{eq:KLR1}) puting aside the instantaneous term:
\begin{equation}
    S_{\text{LF}}(k)= 
\imath\int_{0}^{\infty}ds~\frac{\rho\big(s,\,k_{(s)}^{\text{on}}\big) } {k^2-s+\imath\varepsilon }
~,
\end{equation}
where the on-mass-shell spectral density is
\begin{equation}
\rho(s,\,k_{(s)}^{\text{on}}) = \rho^V (s) \,\slashed{k}_{(s)}^{\text{on}} + \rho^S (s)\, .
\end{equation}

The kernel of the Bethe-Salpeter equation for two quarks propagating generates at zeroth order the free two-body propagator as the disconnected two-quark Green's function propagating in LF time is written as:
\begin{equation}\label{eq:G0D}
\hspace{-0.2 cm} G^\text{LF}_{0} (K)= -\hspace{-.1cm}\int\hspace{-.1cm}d\mu^2 d\nu^2\,\frac{ \rho\big(\mu^2,\,{k}^\text{on}_{1(\mu^2)}\big)\otimes \rho\big(\nu^2,\,{k}^\text{on}_{2(\nu^2)}\big)}{(k_1^2-\mu^2+\imath\varepsilon )\,(k_2^2-\nu^2+\imath\varepsilon )}\,,
\end{equation}
where $\mu^2$ and $\nu^2$ denote the spectral mass parameters of the dressed quark propagators. $G^\text{LF}_{0} (K)$
is interpreted as the probability amplitude for the two dressed quarks propagating independently in their individual LF times.

The disconnected LF Green's function, Eq.~\eqref{eq:G0D}, is rewritten with the LF momenta, namely:
\begin{multline}\label{eq:G0D1}
\hspace{-0.2 cm} G^\text{LF}_{0}(K) = -\int d\mu^2 d\nu^2\,\frac{ \rho\big(\mu^2,\,{k}^\text{on}_{1(\mu^2)}\big)}{k^+_1(k^-_1-{k}^{on,-}_{1(\mu^2)}+\frac{\imath\varepsilon }{k^+_1})}
\\
\otimes
\frac{ \rho\big(\nu^2,\,{k}^\text{on}_{2(\nu^2)}\big)}{k^+_2(k^-_2-{k}^{on,-}_{2(\mu^2)}+\frac{\imath\varepsilon }{k^+_2})}\,,
\end{multline}
where
\begin{equation}
    K^{\pm}=k_1^\pm+ k_2^\pm\quad\text{and}\quad \mathbf{K}_\perp=\mathbf{k}_{1\perp}+\mathbf{k}_{2\perp}\,,
\end{equation}
which expresses the constraint of the total LF momentum conservation. Note that for the sake of simplicity the $G^\text{LF}_{0}(K)$ dependence on $k_1$ is not explicitly exhibited.

The elimination of the relative LF time between quarks 1 and 2, leads to the definition of the resolvent of two dressed quarks derived by integration over $k^-_1$, respecting the momentum conservation~\cite{Marinho:2008pe}. This leads to the LF equal-time (eqt) resolvent:
\begin{equation}\label{eq:G0eqt1}
G^\text{LF}_0(K)\big|_\text{eqt} = \int^\infty_{-\infty} dk_1^{-} \,G^\text{LF}_0(K)\,.
\end{equation}
 Performing the contour integration over the $k^-_1$ lower half-plane, we obtain using the residue theorem:
\begin{equation}\label{eq:G0eqt}
G^\text{LF}_0(K)\big|_\text{eqt} = 2\pi\,\imath\, 
\frac{\theta(k^+_1)}{k^+_1}\frac{\theta(k^+_2)}{k^+_2}\,\, g_0(K^-)\,,
\end{equation}
 where the LF resolvent is:
\begin{equation}\label{eq:g0rho}
g_0(K^-)=\hspace{-.05cm}\int\hspace{-.05cm}d\mu^2 d\nu^2\,\frac{ \rho\big(\mu^2,\,{k}^\text{on}_{1(\mu^2)}\big)\otimes \rho\big(\nu^2,\,{k}^\text{on}_{2(\nu^2)}\big)}{K^- -k^{ \text{on},-}_{1(\mu^2)} -k^{ \text{on},-}_{2(\nu^2)}+\imath\varepsilon }\,.
\end{equation}

The Fourier transform of $G^\text{LF}_0(K)\big|_\text{eqt}$ in $K^-$ provides the forward LF time evolution of the two distinct dressed fermions, but not interacting. In the case the Fourier transform is done in $k^+_1$, $k^+_2$, $\mathbf{k}_{1\perp}$ and $\mathbf{k}_{2\perp}$ it will provide the probability amplitude for the LF-time interval $x^+$ that two fermions  at  given positions  $\{x^-_{1(2)},\,\mathbf{x}_{1(2)\perp}\}$ evolves to other positions onto the null-plane.

As a matter of fact, in the case of two free distinct fermions, when the spectral densities lead to a single time-like pole in the Dirac propagator, the resolvent turns into the standard form:
\begin{equation}
g_0(K^-)=\frac{ \rho\big(m^2,\,{k}^\text{on}_{1(m^2)}\big)\otimes \rho\big(m^2,\,{k}^\text{on}_{2(m^2)}\big)}{K^- -k^{ \text{on},-}_{1(m^2)} -k^{ \text{on},-}_{2(m^2)}+\imath\varepsilon }\,,
\end{equation}
where the denominator can be expressed as
\begin{equation}
K^+\Big( K^- -k^{ \text{on},-}_{1(m^2)} -k^{ \text{on},-}_{2(m^2)} \Big) =    M^2-M_0^2\,,  
\end{equation}
with $M^2=K\cdot K$ and one recognizes the free mass-squared operator as:
\begin{equation}
 M_0^2={\mathbf{k}^2_{1\perp}+m^2 \over x}+{\mathbf{k}^2_{2\perp}+m^2 \over 1-x}-\mathbf{K}_\perp^2 \,,  
\end{equation}
where $x=k^+_1/K^+$ is the momentum fraction. The squared free mass operator is an essential ingredient for phenomenological valence wave function, as well as the kinetic part of the interacting square mass operator in dynamical frameworks, as  Basis Light Front Quantization (BLFQ)~\cite{Vary:2016emi}.

\section{Quark Propagator Model} \label{sec:QuarkProp}

The dressed quark propagator in covariant gauges is written generally in the form:
\begin{equation}
    S(p)
    = i \, Z(p^2)~{\psla p +{\cal M}(p^2)\over p^2-{\cal M}^2(p^2)} 
    ~,
\label{eq:SMZ}
\end{equation}
where ${\cal M}(p^2)$ is the dressed quark-mass  and $Z(p^2)$  the wave-function renormalization.

For our purposes we will use a model for the quark self-energy inspired in lattice QCD calculations.  The adopted phenomenological model was proposed in Ref.~\cite{Mello:2017mor} and  updated in \cite{Castro:2023bij}. The quark-mass function ${\cal M}(p^2)$ was obtained by a fit to Landau gauge  LQCD results for space like-momentum~\cite{Oliveira:2018lln,Oliveira:2020yac,Oliveira:2025boh}.

The fitting formula for the dressed quark mass contains a single  pole in the timelike region
\begin{equation}
    {\cal M}(p^2)=m_0-{m^3\over p^2-\lambda^2+\imath\varepsilon }~ ,
\label{eq:Mk2}
\end{equation}
with parameters from Ref.~\cite{Castro:2023bij} and given by
$m_0=0.008\,\text{GeV}~, \quad m=0.648\,\text{GeV and } \quad  \lambda=0.9\,\text{GeV}$.

This simple  parametrization provides reasonable reproduction of  LQCD calculations and presented in  Ref.~\cite{Oliveira:2018lln}, while the infrared (IR) mass 
$$m_{IR}={\cal M}(0)=m_0+ m^3/\lambda^2={\cal M}^{LQCD}(0)=0.344\,\text{GeV},$$ and $m_0=m^{LQCD}_0=0.008$ GeV
are taken equal to the values of  the accurate parameterization of LQCD calculations proposed in  Ref.~\cite{Oliveira:2020yac}.

The dressed propagator of   Eq.~\eqref{eq:SMZ} has poles for $m_i={\cal M}(m^2_i)$, and the model presents  only three poles from the solutions of the cubic equation: 
\begin{equation}
   m_i(m^2_i-\lambda^2)=\pm[ m_0(m_i^2-\lambda^2)-m^3] \, .
\label{eq:poles}\end{equation}
The spectral densities $\rho^{V(S)}(s)$ in Eq. \eqref{eq:KLR} are  given  by a sum of three Dirac delta-functions \cite{Mello:2017mor}, viz.
\be
    \rho^{S(V)}(s) = \sum_{a=1}^{3} R^{S(V)}_a\delta(s-m_a^2)~, 
\label{eq:rho}
\ee
where
$R^{S(V)}_a$ are the residues,   that read  
\begin{equation}
    R^{V}_a = \frac{(\lambda^2-m^2_a)^2}{(m_a^2-m_b^2)(m^2_a-m^2_c)}~, \,\,
    R^{S}_a = R^{V}_a~{\cal M}(m_a^2),
\label{eq:resid}
\end{equation}
with the indices $\{a,b,c\}$ following the cyclic permutation $\{1,2,3\}$. The poles and residues are given in Table \ref{Tab:polres}. They fulfill   the following relations
\begin{multline}
\sum_{a=1}^3R_a^{V}=1\,,\quad
\sum_{a=1}^3R_a^{S}=m_0\, \,,
     \\    \sum_{a=1}^3m_a^2R_a^{V}=-2\lambda^2+\sum_{a=1}^3 m^2_a\,.
\end{multline}

\begin{table}[htb]
 \caption{Poles, $m_a$, and residues, $R_a$,(cf Eqs. \eqref{eq:rho} and \eqref{eq:resid}) for the fit to the LQCD mass function in Ref. \cite{Oliveira:2018lln}.  The IR mass $m_{IR}=m_0+m^3/\lambda^2$ is 0.344\,GeV, and the parameters of the running mass are also given in the Table. }
  \label{Tab:polres}
\begin{center}
 \begin{tabular}{c c c c}
 \hline  \hline
  ~ $a$ ~ & $m_a$ [GeV] & $R^V_a$ & $R^S_a$[GeV] \\ 
\hline
1  & 0.4695 & 3.7784  & 1.7742\\
2  & 0.5734  & -2.8863 & -1.6546\\
3  & 1.0347  & 0.1079 & -0.1115  \\
 \hline
\hline
&$m_0[GeV]$ & $~~m[GeV]$ & $~~\lambda[GeV]$
\\
\hline
& 0.0080 & 0.6477 &
0.8999\\
\hline
 \end{tabular}
 \end{center}
 \end{table}
We remind that this model violates the positivity relations for the K\"allen-Lehmann  spectral densities of a physical particle, namely
~\cite{itzykson2012quantum}:
\begin{equation}\label{posab}
\rho^V(s)\geq 0 \,\, \,\,\text{and}\,\,\,\,\sqrt{s}\,\rho^V(s)-\rho^S(s)\geq 0 \, .
\end{equation}
However, in QCD the quark  cannot be  an asymptotic state and therefore  the weights in Eq.~\eqref{eq:KLR} should violate the positivity constraints, which is the case in the adopted model.  From Table~\ref{Tab:polres} one has $R^V_2<0$ implying the  violation of the positivity constraint $\rho^V(s)\geq 0$ for $s=m_2^2$. 

\subsection{LF resolvent model}

The light-front propagating part of the fermion two-point function in the model is written taking into account the spectral densities of Eq.~\eqref{eq:rho}, and it is reduced to:
\begin{equation}
       S_{\text{LF}}(k)= 
i\sum_{a=1}^3~\frac{R_a^V\, \psla k_{(m^2_a)}^{\text{on}}+R_a^S} {k^2-m^2_a+\imath\varepsilon } 
\,,\label{eq:slf} 
\end{equation}
where the poles and  residues are given in Table~\ref{Tab:polres}.

The two-quark LF propagator for the running mass model is obtained by introducing the model spectral densities \eqref{eq:rho} in the resolvent Eq.~\eqref{eq:g0rho}, which results in:
\begin{equation}\label{eq:g0}
\hspace{-.1cm}g_0(K^-)=\hspace{-.1cm}\sum_{a,b=1}^3\hspace{-.1cm}
\frac{
\left( R_a^V\sla{k}^\text{on}_{1,a} + R^S_a \right)  \otimes \left( R_b^V\sla{k}^\text{on}_{2,b} + R^S_b \right)}
{K^- - \frac{\mathbf{k}^2_{\perp} + m^2_a}{k^+_1 } 
-  \frac{\mathbf{k}^2_{\perp} + m^2_b}{k^+_2 }
	 +\imath\varepsilon }\,,
\end{equation}
where $k^{\text{on}}_{i,a} \equiv k^{\text{on}}_{i,(m_a^2)}$. The particular frame $\mathbf{ K}_\perp=\mathbf{k}_{1\perp}+\mathbf{k}_{2\perp}=0$ is chosen and $\mathbf{k}_{\perp}\equiv\mathbf{k}_{1\perp}=-\mathbf{k}_{2\perp}$. 
The individual on-shell LF energy follows from Eq.~\eqref{eq:kon}, taken at each mass pole namely, $s=m^2_a$ $(a=1,2,3)$, written as:
\begin{equation}
k^{\text{on},-}_{i,a}\equiv k^{\text{on},-}_{i(m_a^2)} = \frac{(\mathbf{k}_{i\perp}^2 + m_a^2)}{k_{i}^+}\,,
\end{equation}
with $i=1$ or 2 labeling the quarks. 
The resolvent is implicitly a function of $k^\pm_1$ and $\mathbf{k}_{1\perp}$, with the kinematical momentum of fermion~2 constrained by total momentum conservation. 

The Fourier transform in $K^-$ give the disconnected two-point function, which corresponds to  a pair of dressed  quarks propagating  forward in the common LF time. It defines the ``free" LF resolvent for two dressed fermions with no interactions among them.

\section{Dressed mass operator} \label{sec:DressedMO}
The  eigenvalue equation for the interacting mass-squared operator for the valence component of the quark-antiquark bound state used in BLFQ or obtained by the LF projection of the Bethe-Salpeter equation~\cite{Frederico:2010zh} is:
\begin{equation}\label{eq:masssquareoperator}
\left( M^2_0 + M^2_I \right) | \Psi_{q \bar{q}} \rangle = K^2| \Psi_{q \bar{q}}\rangle\,,
\end{equation}
where $M^2_0$ is the quark-antiquark free mass-squared operator and $M^2_I$ is the effective interaction. For dressed quarks the free mass operator has to be redefined considering the resolvent $g_0(K^-)$ given by Eq.~\eqref{eq:g0}. 

The idea is to project the $g_0(K^-)$  in a constituent quark spinor basis defined by the infrared mass, which allow to introduce the physics of the   dressed quarks in the mass-squared eigenvalue equation, while keeping the practical form expressed by Eq.~\eqref{eq:masssquareoperator}.  Haag's theorem poses the challenge that the Hilbert space of the free and interacting QFT's are different, which we face by taking a practical route to project the $g_0(K^-)$ in a basis of quarks with a fixed mass. Such approach is developed in what follows.

We introduce the on-mass-shell momentum $k^{IR}$ such that $k^{IR}\cdot k^{IR}=m_{IR} ^2$ holds, with $m_{IR}$ defined by the running mass~\eqref{eq:Mk2} at $p^2=0$. The positive energy spinor projector is defined according to
\begin{equation}
	P^{IR}_1=\frac{ \left( \sla{k}^{IR}_{1} + m_{IR} \right)}{2 m_{IR}}\,,
\end{equation}
for the quark free states with fixed infrared mass. Then, the resolvent $g_0$ from Eq.~\eqref{eq:g0} is projected in these states by using $P^{IR}_1$ and $P^{IR}_2$, which allows to define an invertible operator 
\begin{equation}\label{eq:tildeg0}
\tilde g_{0,m_{IR}} {(K^-)}=P^{IR}_1\otimes P^{IR}_2\,g_0 (K^{-})\,P^{IR}_1\otimes P^{IR}_2 \,,
\end{equation}
and from that an effective mass-squared operator, $M_{0,D}^2$ (see Eq. \eqref{eq:M0D}), which takes into account the effect of the quark dressing. 

However, an intrinsic dependence on the eigenvalue $K^2=K^-K^+-\mathbf K_\perp^2$ is carried by \eqref{eq:tildeg0}, which forbids the formulation of the mass-squared eigenvalue equation in a simplistic form. For that reason, for now on we will assume that $K^2<<m_i^2$ and  $K^2<<m_{IR}^2$, addressing particular
applications such as the pion physics, which in the chiral limit has a vanishing mass. Considering such an illustrative case we define:

\begin{equation}\label{eq:M0D}
\hspace{-0.2cm} M_{0,D}^2\,P^{IR}_1\otimes P^{IR}_2= Z_M\,\lim_{K^2 \to 0}~K^+ \,  [\tilde g_{0,m_{IR}}(K^-) ]^{-1} \,.
\end{equation}
We observe that the product $K^+ \,  [\tilde g_{0,m_{IR}}(K^-) ]^{-1}$ depends on $K^2 =K^-K^+-\mathbf K_\perp^2$. $Z_M$ is a renormalization factor to enforce the physical condition that in the UV limit the physical dressed mass operator,
\begin{equation}\label{eq:m02phys}
M^2_{0,D}|_\text{phys}\equiv M_{0,D}^2- \frac{C}{x_1x_2} \,,
\end{equation}
meets the free one, namely:
\begin{small}
\begin{equation}\label{eq:m02physlim}
\hspace{-.1cm}\lim_{\mathbf{k}^2_{\perp}\to\infty}\hspace{-.1cm}M^2_{0,D}|_\text{phys}\to\frac{k^2_\perp+m_0^2}{x_1x_2}  \,,\end{equation}\end{small}
where the momentum fractions are $x_{1}=k^+_1/K^+$ and $x_{2}=k^+_2/K^+$, with $x_1+x_2=1$ and $x_1\equiv x$.
The subtraction constant $C$ assures that for large $|\mathbf{k}_\perp|$ the physical dressed square mass operator tends to the form shown in Eq.~\eqref{eq:m02phys} with $m_0$ being the bare quark mass.  Note that the inverse operator $~\tilde g_{0,m_{IR}}^{-1}$ is meaningful in the subspace of positive energy spinors.  The renormalization constant $Z_M$ is derived in what follows  and $C$ will be  obtained in Sect.~\ref{sec:ELCQSE}.

Our task now is to evaluate the operator
\begin{equation}
A_{1,a}= P^{IR}_1
	  \left( R^V_a\sla{k}^\text{on}_{1,a}+R^S_a \right)
	   P^{IR}_1 .\label{opA}
\end{equation}

To derive an explicit form for this operator we can start by considering the group algebra for the Dirac matrices
$\{  \gamma^\mu,\gamma^\nu \}  \ =  \ 2 g^{\mu \nu} $,  which leads to: 
$$ \sla{k}^{IR}_1 \sla{k}^\text{on}_{1,a} = - \sla{k}^\text{on}_{1,a} \sla{k}^{IR}_1 + 2 k^\text{on}_{1,a} \cdot k^{IR}_1\, ,$$ 
and introducing this identity into Eq.~\eqref{opA} we factor out the spinor projection operator, viz, 
\begin{equation}
\label{eq:Ai}
A_{1,a}= \zeta_a(k_{1,a}^\text{on}\cdot k^{IR}_1)\,\frac{\left( \sla{k}^{IR}_1  + m_{IR}   \right) }{2 m_{IR} } \,,
\end{equation}
where the effective fermion operator $\zeta_a$ carries the quark dressing from the Dirac algebra:
\begin{equation}
 \zeta_a(y ) = R^S_a+ \,\frac{R^V_a}{m_{IR}} \,y\,.
\end{equation}
Thus, taking the operators $A_{1a}$ and the analogous for $A_{2a}$ in the form  of Eq.~\eqref{eq:Ai} into Eq.~\eqref{eq:tildeg0}, we can write the effective resolvent as:
\begin{multline}
    \tilde g_{0,m_{IR}} (K^-)=\sum_{a,b=1}^3 \frac{A_{1,a}\otimes  A_{1,b}}{K^- - \frac{\mathbf{k}^2_{1\perp} + m^2_a}{k^+_1 } 
-  \frac{\mathbf{k}^2_{2\perp} + m^2_b}{k^+_2 }
	 +\imath\varepsilon }= \\
     \hspace{-.2cm}P^{IR}_1\otimes P^{IR}_2
     \sum_{a,b=1}^3 \frac{\zeta_a(k_{1,a}^\text{on}\cdot k^{IR}_1)\,\zeta_b(k_{2,b}^\text{on}\cdot k^{IR}_2)}{K^- - \frac{\mathbf{k}^2_{1\perp} + m^2_a}{k^+_1 } 
-  \frac{\mathbf{k}^2_{2\perp} + m^2_b}{k^+_2 }
	 +\imath\varepsilon }\, .
\end{multline}

Now this is our effective resolvent in a good approximation for the resolvent of quark-antiquark systems in LF dynamics, and we will proceed to analyze the mass-squared operator for hadron physics along with some other phenomenological observables starting from this expression.

The effective dressed mass-squared operator defined in Eq.~\eqref{eq:M0D} is given by:
\begin{equation}\label{eq:M0DC}
 \hspace{-.15cm}   M^2_{0,D}=Z_M\,\left[ \sum_{a,b=1}^3\frac{\zeta_a(k_{1,a}^\text{on}\cdot k^{IR}_1)\,\zeta_b(k_{2,b}^\text{on}\cdot k^{IR}_2)}{ \frac{\mathbf{k}^2_{1\perp } + m^2_a}{x_1 } 
+\frac{\mathbf{k}^2_{2\perp } + m^2_b}{x_2 }}\right]^{-1} \,,
\end{equation}
and in the next we determine $Z_M$.

\begin{figure}[t!]
    \centering
\includegraphics[width=1\linewidth]{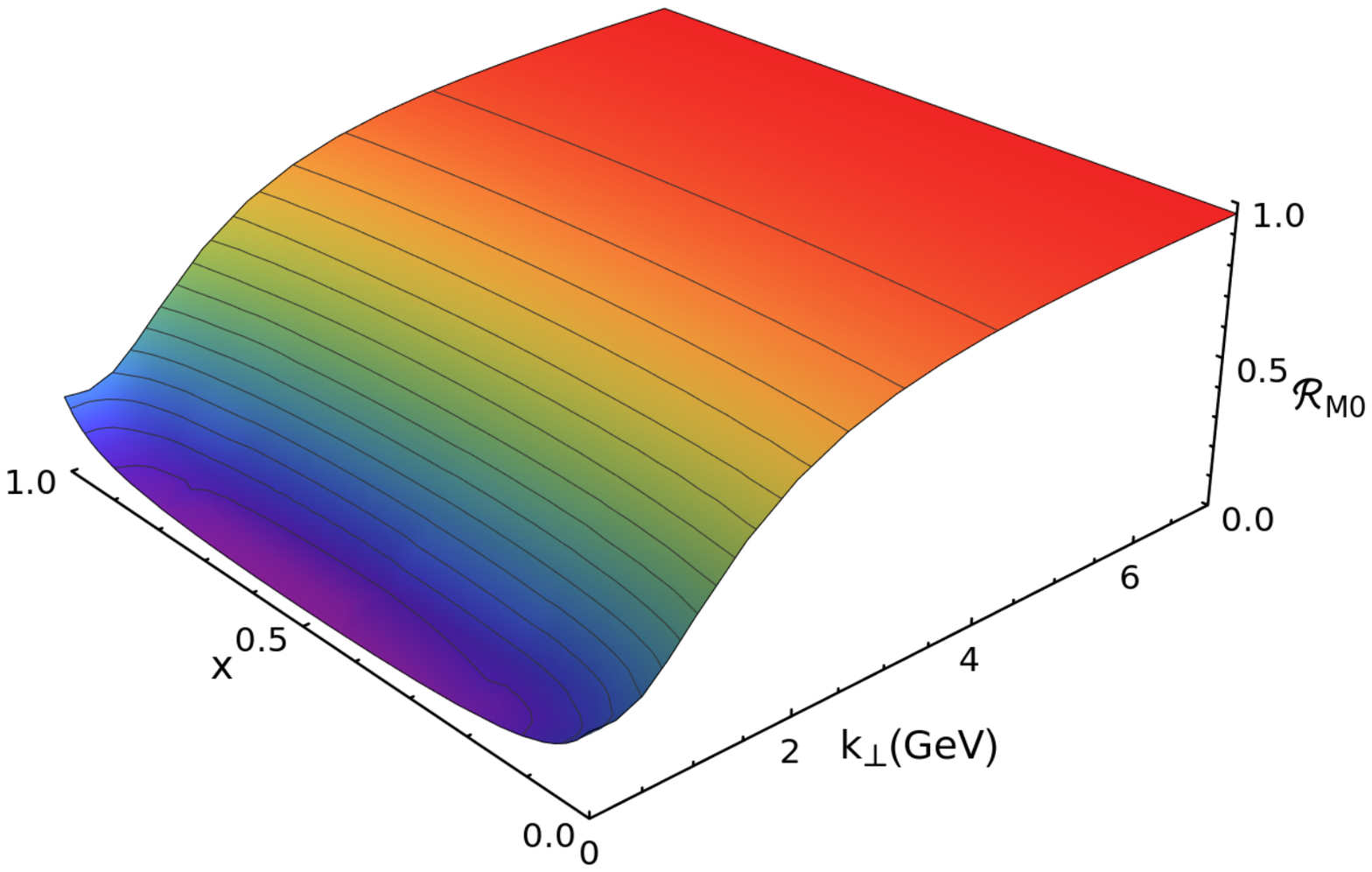}
    \caption{Ratio $ \mathcal{R}_{M_0}={M^2_{0,D}(\mathbf k_\perp^2,x)/M^2_{0,IR}(\mathbf k_\perp^2,x)}\,$ as a function of the momentum fraction $x$ and transverse momentum $k_\perp=|\mathbf{k}_\perp|$. }
    \label{fig:ratio}
\end{figure}

In the UV limit, $|\mathbf{k}_{\perp}|  \to \infty$, the renormalization factor $Z_M$ should be found by imposing that 
$$M^2_{0,D}|_{UV}\to \frac{\mathbf{k}_\perp^2}{x(1-x)}\,,$$ and in this limit Eq.~\eqref{eq:M0DC} is given by:
\begin{eqnarray}
M^2_{0,D}\to Z_M\frac{\mathbf{k}_\perp^2}{x(1-x)}\,\Bigg(\sum_{a=1}^3 \zeta_a( k_{1,a}\cdot  k^{IR}_1)\big|_{UV} \Bigg)^{-2} \!\!\! , 
\end{eqnarray} 
that resolves the renormalization factor as given by:
\begin{eqnarray}
Z_M=\Big(\sum_{a=1}^3 \zeta_a( k_{1,a}\cdot  k^{IR}_1)\big|_{UV} \Big)^2 \, .
\end{eqnarray}
Taking into account the  on-mass-shell momenta, the momentum dependence cancels out, according to:
\begin{equation}
k_{1,a}\cdot  k^{IR}_1=   k^+_{1}\,\frac{  k^{IR-}_1+ k^{on-}_{1,a}} {2} 
- \mathbf{k}^2_{1 \perp}=\frac{m^2_{IR}+m^2_a}{2}\,,
\end{equation}
and the renormalization factor is reduced to
\begin{equation}\label{eq:Z}
Z_M^\frac12=\sum\limits_{a=1}^3\overline \zeta_a\,,\end{equation}
where
\begin{equation}\label{eq:zetaa}
\overline\zeta_a=   R^S_a+R^V_a \,\frac{ m^2_{IR} + m^2_a}{2\,m_{IR}}\,.
\end{equation}

In order to appreciate the effect of the running quark mass in $M^2_{0,D}$, we  compute the ratio with the standard free squared mass of the composite system with constituents having mass $m_{IR}$, given by:
\begin{equation}
    \mathcal{R}_{M_0}={M^2_{0,D}(\mathbf k_\perp^2,x)\over M^2_{0,IR}(\mathbf k_\perp^2,x)}\,,
\end{equation}
where 
\begin{equation}\label{eq:M20IR}
    M^2_{0,IR}(\mathbf{k}^2_\perp,x)=\frac{\mathbf k^2_\perp+m^2_{IR}}{x(1-x)}\,.
\end{equation}
The  ratio $\mathcal{R}_{M_0}$ is shown in Fig.~\eqref{fig:ratio} and a large reduction is seen at low transverse-momentum below 2\,GeV, namely in the IR region where it is expected a strong effect of the QCD interaction. 

\section{Effective LC quark self-energy}\label{sec:ELCQSE}

Now we should compute the constant $C$ in the model.
In order to obtain this constant we introduce an effective quark self-energy $\Sigma_\text{eff}(\mathbf{k}_\perp^2,x)$  within the composite system  to account for the behaviour of the dressing of the mass-squared operator, as:
\begin{equation}
    M^2_{0,D}(\mathbf k_\perp^2,x)|_\text{phys}\equiv \frac{\mathbf k_\perp^2+ \Sigma^2_\text{eff}(\mathbf{k}_\perp^2,x)}{x(1-x)}\,. \label{eq:M20Dphys}
\end{equation}
Manipulating it by introducing $M^2_{0,D}$ from Eq.~\eqref{eq:M0DC}  in Eq.~\eqref{eq:m02phys}, we find that: 
\begin{equation}
    \Sigma^2_\text{eff}(\mathbf{k}_\perp^2,x)=x(1-x) M^2_{0,D}(\mathbf k_\perp^2,x) - \mathbf{k}_\perp^2-C\,,\label{eq:sigma2}
\end{equation}
with the physical condition 
\begin{equation}
\lim_{\mathbf{k}_\perp^2\to\infty}  \Sigma^2_\text{eff}(\mathbf{k}_\perp^2,x)=m_0^2\, . 
\end{equation}
The constant $C$ is derived in App.~\ref{app:constantC} that results in:
\begin{equation}\label{eq:Cfinal}
C= Z_M^{-\frac12} \sum_{a=1}^3  \overline \zeta_a\,m^2_a -m^2_0\,.
\end{equation}
 Once $C$ was obtained, the effective LF quark self-energy, $\Sigma_\text{eff}(\mathbf{k}_\perp^2,x)$ can be computed and the result is:
\begin{multline}\label{eq:qselfen}
    \Sigma^2_\text{eff}(\mathbf{k}_\perp^2,x)=x(1-x) M^2_{0,D}(\mathbf k_\perp^2,x)- \mathbf{k}_\perp^2 \\ 
    - Z_M^{-\frac12} \sum_{a=1}^3  \overline \zeta_a\,m^2_a +m^2_0\,,
\end{multline}
which tends to the current quark mass for large transverse momentum. We should remark that the effective light-front quark self-energy introduced here is an operator-level quantity and not a one-particle–irreducible self-energy computed with  a Schwinger–Dyson equation.
 
\begin{figure}[h]
    \centering
    \includegraphics[width=1\linewidth]{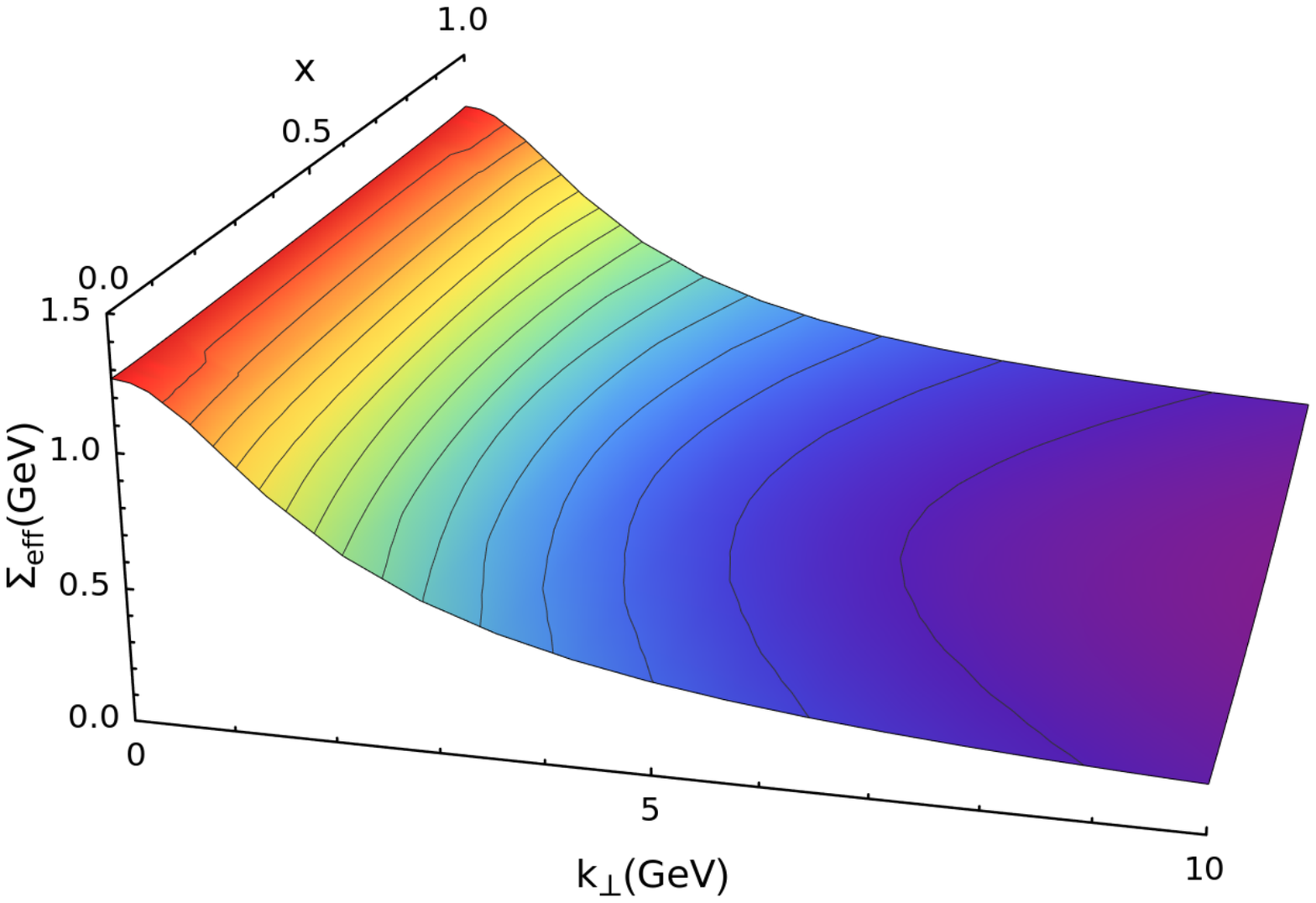}
    \caption{Effective quark self-energy as a function of the momentum fraction $x$ and transverse momentum $k_\perp=|\mathbf{k}_\perp|$.}
    \label{fig:self}
\end{figure}

The result for $\Sigma_\text{eff}(\mathbf{k}_\perp^2,x)$ is shown in Fig.~\ref{fig:self}, where one observe the IR enhancement of the LF quark self-energy. The effective mass of the quarks at small relative momentum grows up to 1.25\,GeV,  which is a couple of times the standard constituent quark mass and it has a scale of about 2\,GeV for the dependence with $k_\perp$. Such a scale seems to  be surprisingly large if compared to $\Lambda_{QCD}\sim 0.3\,$GeV, however the model  quark self-energy has a scale $\lambda=0.9$\,GeV, that drives the running of the LF quark mass. { Furthermore, we observe a weak dependence of the LF effective quark mass on the momentum fraction, which is due to the construction seen in Eq.~\eqref{eq:qselfen}, where the factor $x(1-x)$ is essentially canceled out by the product with $M_{0,D}^2$ in Eq.~\eqref{eq:qselfen}, although it is expected that towards the end-points the UV limit is approached and the self-energy would tend to the current quark mass. Such limitation is due to the projection  done of the resolvent in the quark spinors with a fixed IR mass. }

\begin{figure}[h]
    \centering   \includegraphics[width=1\linewidth]{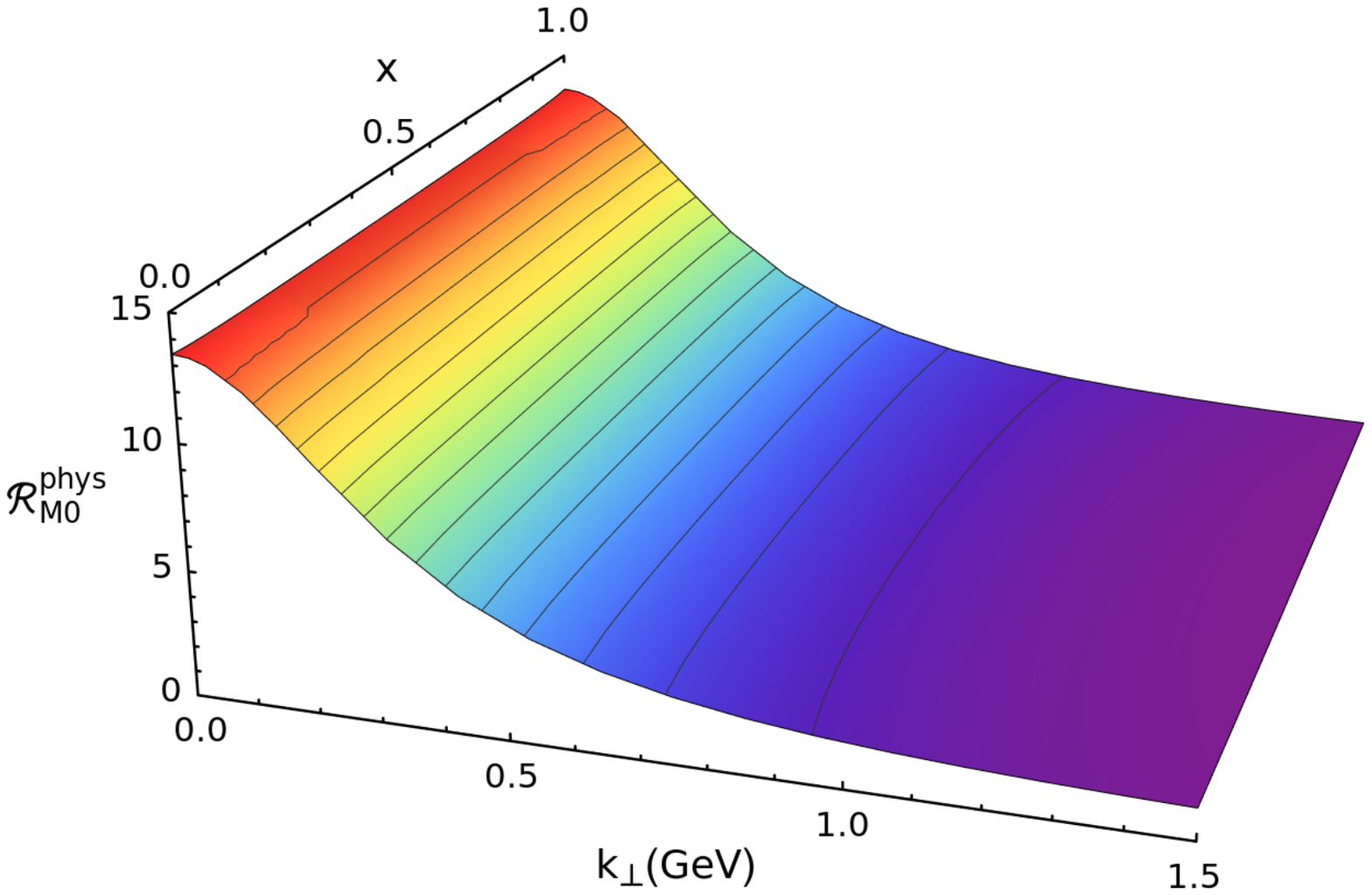}
    \caption{Ratio $ \mathcal{R}^\text{phys}_{M_0}={M^2_{0,D}(\mathbf k_\perp^2,x)|_\text{phys}/M^2_{0,IR}(\mathbf k_\perp^2,x)}\,$ as a function of the momentum fraction $x$ and transverse momentum $k_\perp=|\mathbf{k}_\perp|$.}
    \label{fig:ratiophys}
\end{figure}

The physical free-mass operator with dressed quarks is written in terms of the LF self-energy as provided Eq.~\eqref{eq:M20Dphys},  and computing the ratio
\begin{equation}
\mathcal{R}^\text{phys}_{M_0}={M^2_{0,D}(\mathbf k_\perp^2,x)|_\text{phys}\over M^2_{0,IR}(\mathbf k_\perp^2,x)}\,,
\end{equation}
one has an insight in which regions of the $\{k_\perp,x\}$ plane the effect of the running LF quark mass is relevant within the pion state. 
In Fig.~\ref{fig:ratiophys} the physical ratio, $\mathcal{R}^\text{phys}_{M_0}$, is shown and we observe a huge enhancement in the IR region, which  is expected as a consequence of the QCD interaction be stronger at low momentum or at large distances. The physical mass operator with running effective LF self-energy is not enough to produce  meson eigenstates, as we have started with the disconnected two-quark propagator, or equivalently the quark-antiquark propagator. Other correlations between the $q$  and $\overline q$ have to be introduced in the mass-squared eigenvalue equation as the interaction between these dressed degrees of freedom  to enforces absolute confinement. Furthermore, we observe the large enhancement of $ M^2_{0,D}(\mathbf k_\perp^2,x)|_\text{phys}$, of at least one order of magnitude in the IR region as compared to the constituent $M^2_{0,IR}(\mathbf{k}^2_\perp,x)$ \eqref{eq:M20IR}. Naively, such a large growing of  $ M^2_{0,D}(\mathbf k_\perp^2,x)|_\text{phys}$ favors the quark confinement, coherently to what is expected in QCD at long distances.

\section{Pion Phenomenology} \label{sec:PionPheno}

The squared free-mass operator for a quark-antiquark pair $(M_0^2)$ has been used as an essential ingredient in building phenomenological models of the pion valence light-front wave function, which allows to build simple functions like exponential and polynomial, with $M_0^2$ in  the argument,  motivated by the free  light-cone resolvent form that is written in terms of $\sim 1/(M^2-M_0^2)$ (see ~\cite{BRODSKYPREP}). This dependence  for decreasing function naturally allows to build valence wave functions with the desired property to vanish at the end point and at large transverse momentum. However, such mass operator accounts only for fixed mass constituents, not allowing the freedom to run with LF momentum. That has been worked out in the previous section, where we have derived an effective squared dressed mass operator for a quark-antiquark system  given by Eq.~\eqref{eq:M20Dphys}, which included the effective quark-self energy~(see Eq.~\eqref{eq:sigma2} and Fig.~\ref{fig:self}). At low momentum the effective quark-self energy is large and about 1.2\,GeV and runs to the current quark at large momentum. Such an effective quark self energy, could in principle jeopardize the sucess in describing phenomenology, like for the pion, where the constituent quarks, has a mass around 0.2-0.3\,GeV. Therefore, to check how the valence  models behave when such an effective squared mass for the quark-antiquark system is considered in the arguments of different function, we study the different momentum distributions obtained from Gaussian and polynomials models, and compared to their standard form written as function of the free-mass squared operator with fixed quark mass.

\subsection{Valence models}

Here we consider only the dominant spin antialigned  component of the valence pion wave function. We base our study on  the findings of Ref.~\cite{dePaula:2020qna} where the  Minkowski Bethe-Salpeter equation was solved for the pion, with a kernel including massive gluons and constituent quarks that resulted in a valence probability of 70\% dominated by the antialigned component with 57\%, once the pion mass and decay constant were fitted. Therefore, the models explored for the pion spin antialigned wave function for our study were chosen as:

(A) Gaussian Model:
\begin{equation}\label{eq:expphys}
\Psi^{(A)}_{\uparrow\downarrow}(k_\perp,x) =
\exp\left[ - \frac{1}{m_s^2} \,\frac{k_\perp^2 +  \Sigma^2_\text{eff}(\mathbf{k}_\perp^2,x)}{x (1 - x)} \right] \, ,
\end{equation}
where $ \Sigma^2_\text{eff}(\mathbf{k}_\perp^2,x)$ is defined in Eq.~\eqref{eq:qselfen}. The scale of the momentum dependence of the wave function is $m_s$, which will be varied in our discussion.

(B) The Brodsky-Tao-Lepage wave function~\cite{Huang:1994dy}:
\begin{equation}
\Psi^{(B)}_{\uparrow\downarrow}(k_\perp, x) =
\exp\left[
 - \frac{1}{m_s^2} \,\frac{k_\perp^2 + m_q^2}{x (1 - x)}
\right]\,. \label{eq:expBTL}
\end{equation}

(C) Symmetric vertex  model inspired in Ref.~\cite{Fanelli:2016aqc}:
\begin{multline}\label{eq:psisym}
\Psi^{(C)}_{\uparrow\downarrow}(k_\perp, x)=\frac{1}{x(1-x)}
\frac{1}{
m^2_\pi - \frac{k_\perp^2 + m_{IR}^2}{x (1-x)}
} \\ \times\left[
\frac{1}{
x \left( m^2_\pi - \frac{k_\perp^2}{x(1-x)} - \frac{m_{IR}^2}{1-x} - \frac{m_s^2}{x} \right)
} \right. \\
\times \left. \frac{1}{1-x} \frac{1}{
m^2_\pi - \frac{k_\perp^2}{x(1-x)} - \frac{m_{IR}^2}{x} - \frac{m_s^2}{1-x}
}\,
\right] \, ,
\end{multline}
where the first factor $\frac{1}{x(1-x)}$ in Eq.~\eqref{eq:psisym} is introduced to reproduce the expression of $f_\pi$ in the limit of $m_s\to \infty$ apart a normalization factor derived in Ref.~\cite{FredPRD1992}.

 (D) Symmetric vertex model including the running quark mass from Eq.~\eqref{eq:qselfen}:
\begin{multline}
\Psi^{(D)}_{\uparrow\downarrow}(k_\perp, x) = \frac{1}{x(1-x)}\frac{1}{
m^2_\pi - \frac{\mathbf{k}_\perp^2 +\Sigma^2_\text{eff}(\mathbf{k}^2_\perp,x)}{x (1-x)}
}\\ \times
\left[
\frac{1}{
x \left( m^2_\pi - \frac{k_\perp^2}{x(1-x)} - \frac{m_{IR}^2}{1-x} - \frac{m_s^2}{x} \right)
} \right. \\
\times \left. \frac{1}{1-x} \frac{1}{
m^2_\pi - \frac{k_\perp^2}{x(1-x)} - \frac{m_{IR}^2}{x} - \frac{m_s^2}{1-x}
}
\right]\,.\label{eq:psisym1}
\end{multline}

\begin{table}[t]
    \centering
    \begin{tabular}{|c|c|c|c|c|}
\hline
     Model    &~~ (A) ~~&~~ (B) ~~& ~~(C) ~~& ~~(D) ~~ \\
     \hline \hline
      $m_s$ [GeV]    & 0.58 & 0.73 & 0.5 & 0.175  \\ \hline  
    \end{tabular}
    \caption{Values of the mass scale parameter $m_s$ for the model valence wave functions for (A) Eq.~\eqref{eq:expphys},(B) Eq.~\eqref{eq:expBTL}, (C) Eq.~\eqref{eq:psisym} and (D) Eq.~\eqref{eq:psisym1}, obtained by fixing $f_\pi=130\,$MeV and $P_{q\bar q}=$70\%. }
    \label{tab:msvalues}
\end{table}

The consequences of the dressed $q\bar q$ square mass operator derived in Sect.~\ref{sec:ELCQSE} for the pion properties will be explored based on the definitions given in~\cite{dePaula:2020qna} for the pion decay constant:
\begin{equation}
    f_\pi={N_c\over 2\pi^2}\int^\infty_0dk_\perp k_\perp\int^1_0 dx\, \Psi_{\uparrow\downarrow}(k_\perp,x)\,,
\end{equation}
where $N_c=3$ is the number of colors, 
and the valence probability for the spin antialigned state can be written as:
\begin{equation}
    P_{q\bar q}={N_c \over 4 \pi^2}\int^\infty_0dk_\perp k_\perp\int^1_0 dx\, |\Psi_{\uparrow\downarrow}(k_\perp,x)|^2\,,
\end{equation}
assuming the dominance of the antialigned $q\bar q$ spin configuration. Here we will make use of the experimental value of $f_\pi=130\,$MeV~\cite{PDG2024_PseudoscalarMesonDecays} to fix the normalization of the valence state, leading to a probability of this component in the pion. Then, by giving the valence probability the mass scale parameter can be computed for each of the models from (A) to (D). 

In Table~\ref{tab:msvalues} it is presented the mass scale parameter for a probability of the antialigned component of 70\% and $f_\pi=130\,$MeV. The mass scale parameters ranges from 0.175 to 0.72\,GeV, for  the antialigned spin wave functions. The Gaussian models are less sensitive to the effective LF quark self-energy, as expressed by the closer values of $m_s$, 0.59 and 0.72\,GeV, while the symmetric vertex models, with a power-law like form presents values of $m_s$ of 0.5 and 0.175\,GeV, for models (C) and (D) respectively. Such quite different values of the mass scale parameter compensate the  large values attained by $\Sigma_\text{eff}$ at low transverse momentum in model (D) in comparison to (C), where all mass scales are similar.

\subsection{Momentum distributions}

The phenomenological exploration of the pion structure is illustrated in what follows through the dominant antialigned spin component of the valence state given by the models (A)-(D), where the corresponding  mass scale parameters (see Table~\ref{tab:msvalues}) were obtained by using the experimental value of $f_\pi$ and assuming 70\% for the valence probability. We choose to illustrate the effect of the effective LF quark self-energy in the squared mass operator for the pion state represented by Gaussian  and polynomial-like models of the valence wave function to identify possible sharp differences that may occur due to the incorporation of the running quark mass, and the physics associated with spontaneous chiral symmetry breaking in LF dynamics implicitly carried out by $\Sigma_\text{eff}(\mathbf{k}_\perp^2,x)$ within the pion state. For this purpose, we are going to compare the twist-2 unpolarized Transverse Momentum Distribution (uTMD) $f_1\left(x, k_{\perp}\right)$, the unpolarized Parton Distribution Function (uPDF) $u(x)$ and the Distribution Amplitude (DA) $\phi(x)$ of the different models.

The twist-2 uTMD, the uPDF and the DA for the valence state of the pion, are given, respectively, by:
\begin{eqnarray}
f_1\left(k_{\perp},x\right) &=&\left|\Psi_{\uparrow\downarrow}\left( k_{\perp},x\right)\right|^2 \,, \\
u(x) &=& \int d^2 k_{\perp} f_1\left( k_{\perp},x\right)\,, \\
\phi(x)&=& \int d^2k_\perp \Psi_{\uparrow\downarrow}(k_\perp,x) \, .
\end{eqnarray}

\begin{figure}[h]
    \centering

\includegraphics[width=0.8\linewidth]{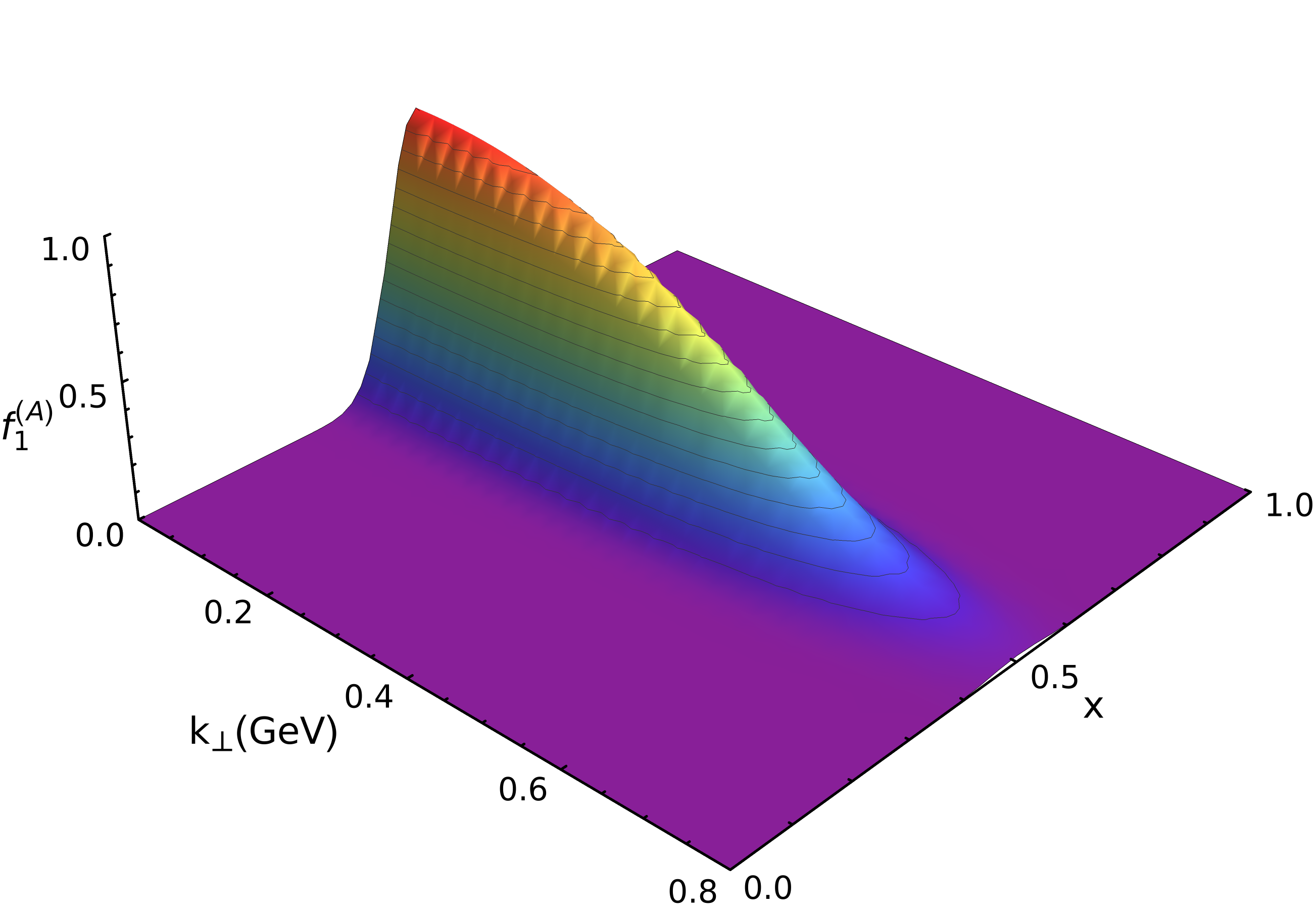}
    
    \includegraphics[width=0.8\linewidth]{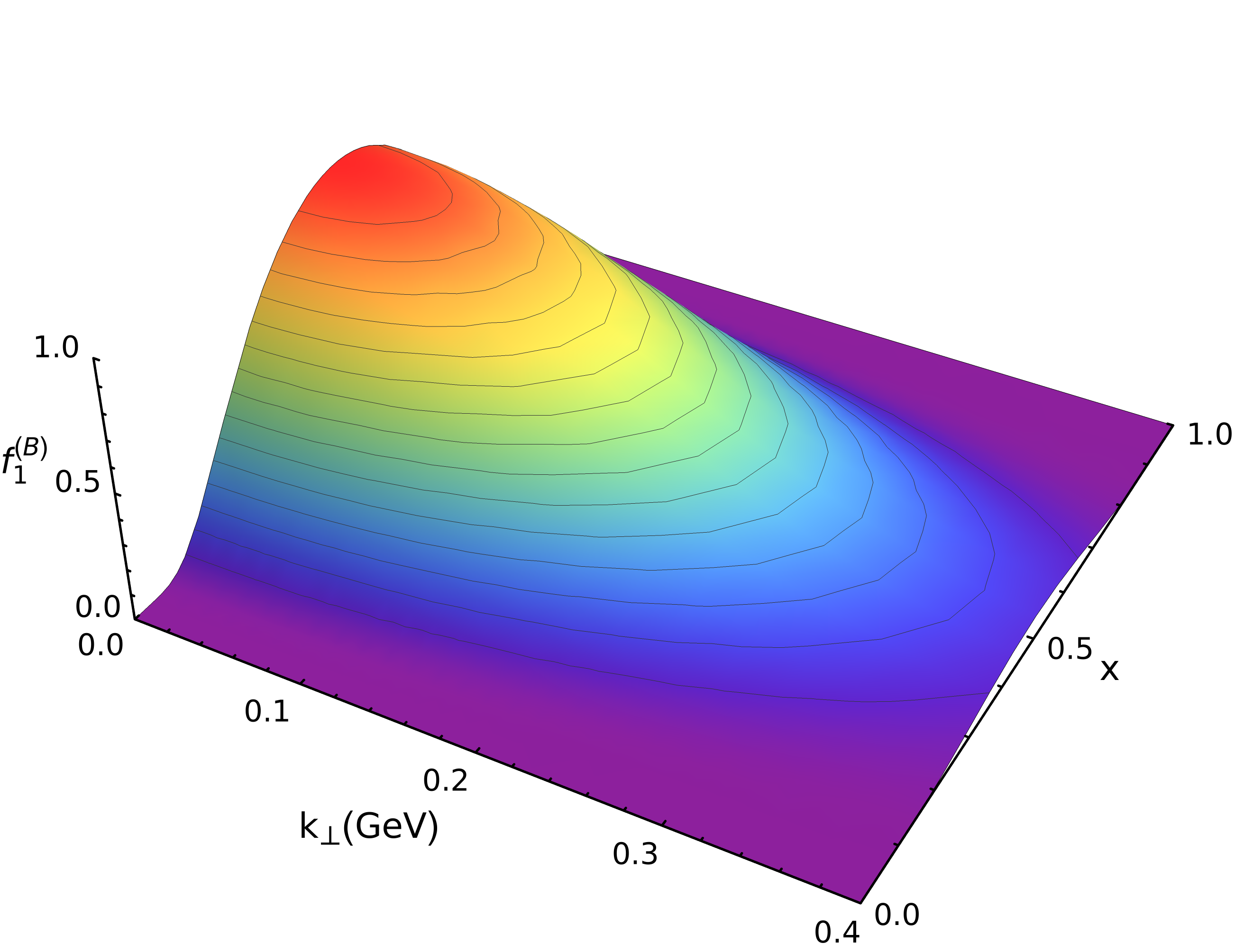}

    \caption{Unpolarized Transverse Momentum Distribution (arbitrary normalization), $f_1$, for the models (A) and (B) parametrized  with $f_\pi=130\,$MeV and  $P_{q\bar q}=70\%$. 
     Top: valence wave function (A) from Eq.~\eqref{eq:expphys} with $m_s=0.58\,$GeV. Bottom: running quark LF self-energy and $m_{IR}$ in the vertex function in the wave function from Eq.~\eqref{eq:expBTL}  model (B) with constituent quark mass of $m_{IR}=0.344\,$GeV and mass scale $m_s=0.73\,$GeV. 
    }
    \label{fig:f1_AB_2D}
\end{figure}

In Fig. \ref{fig:f1_AB_2D} it is shown the unpolarized Transverse Momentum Distribution (arbitrary normalization), $f_1\left(k_{\perp},x\right)$, for the models (A) and (B) parametrized  with $f_\pi=130\,$MeV and  $P_{q\bar q}=70\%$. 
     In the top panel, the uTMD for model A from Eq.~\eqref{eq:expphys} is parametrized with $m_s=0.58\,$GeV for $70\%$ of valence content of the pion state, and the running mass $\Sigma_\text{eff}(\mathbf{k}_\perp^2,x)$ presents at low transverse momentum value of about $1.3$\,GeV, 
     decreasing with $k_{\perp}$ with a characteristic scale of $\sim 2$\,GeV, which makes this model similar of a heavy quark system and therefore for $x=0.5$ the uTMD has a narrow peak. On the other hand, in the bottom panel, that shows the uTMD for model B, has a fixed constituent quark mass of $m_q = m_{IR}=0.344\,$GeV which presents the wide peak,  characteristic of the wave function of a light quark system.

\begin{figure}[h]
    \centering

\includegraphics[width=0.8\linewidth]{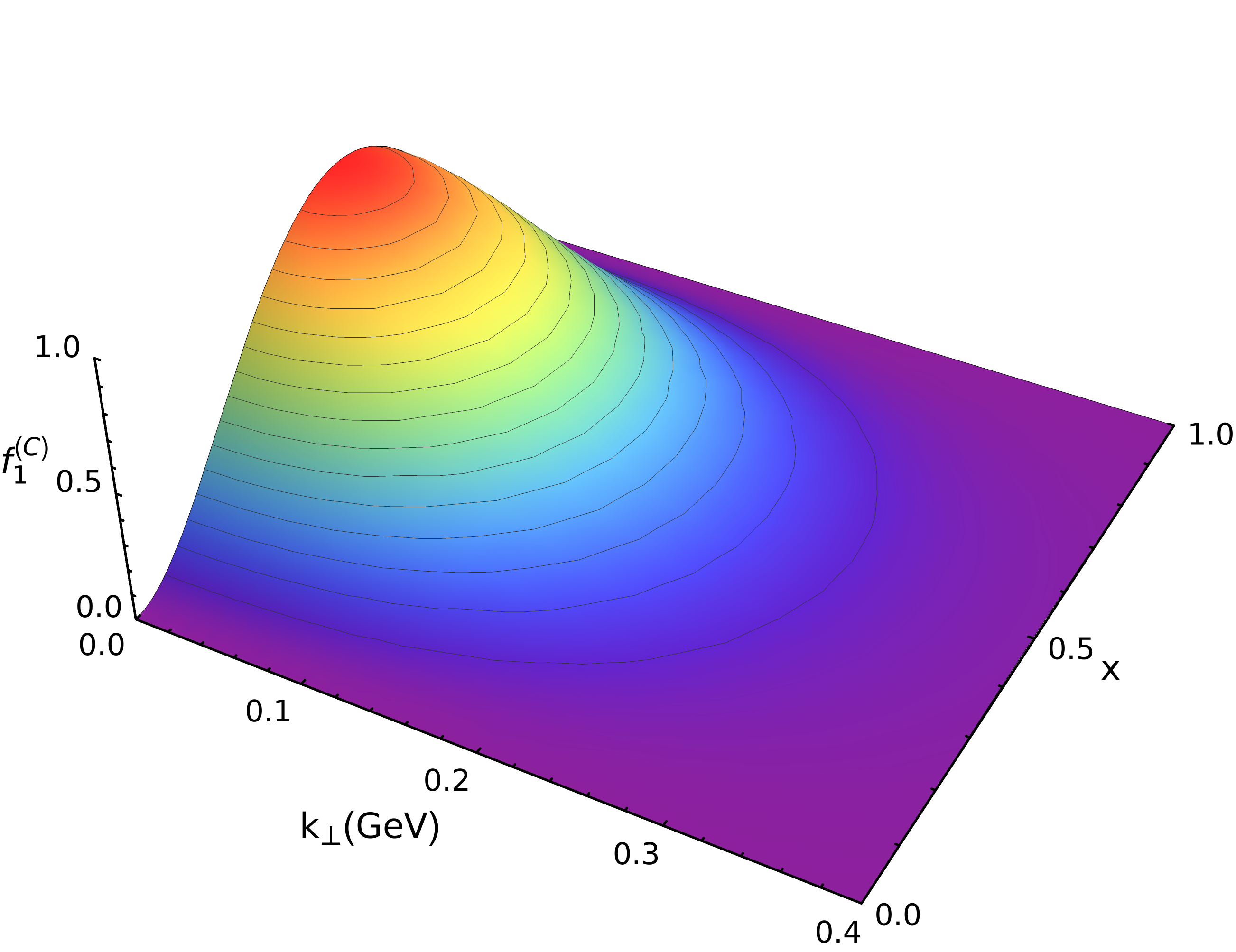}
    
    \includegraphics[width=0.8\linewidth]{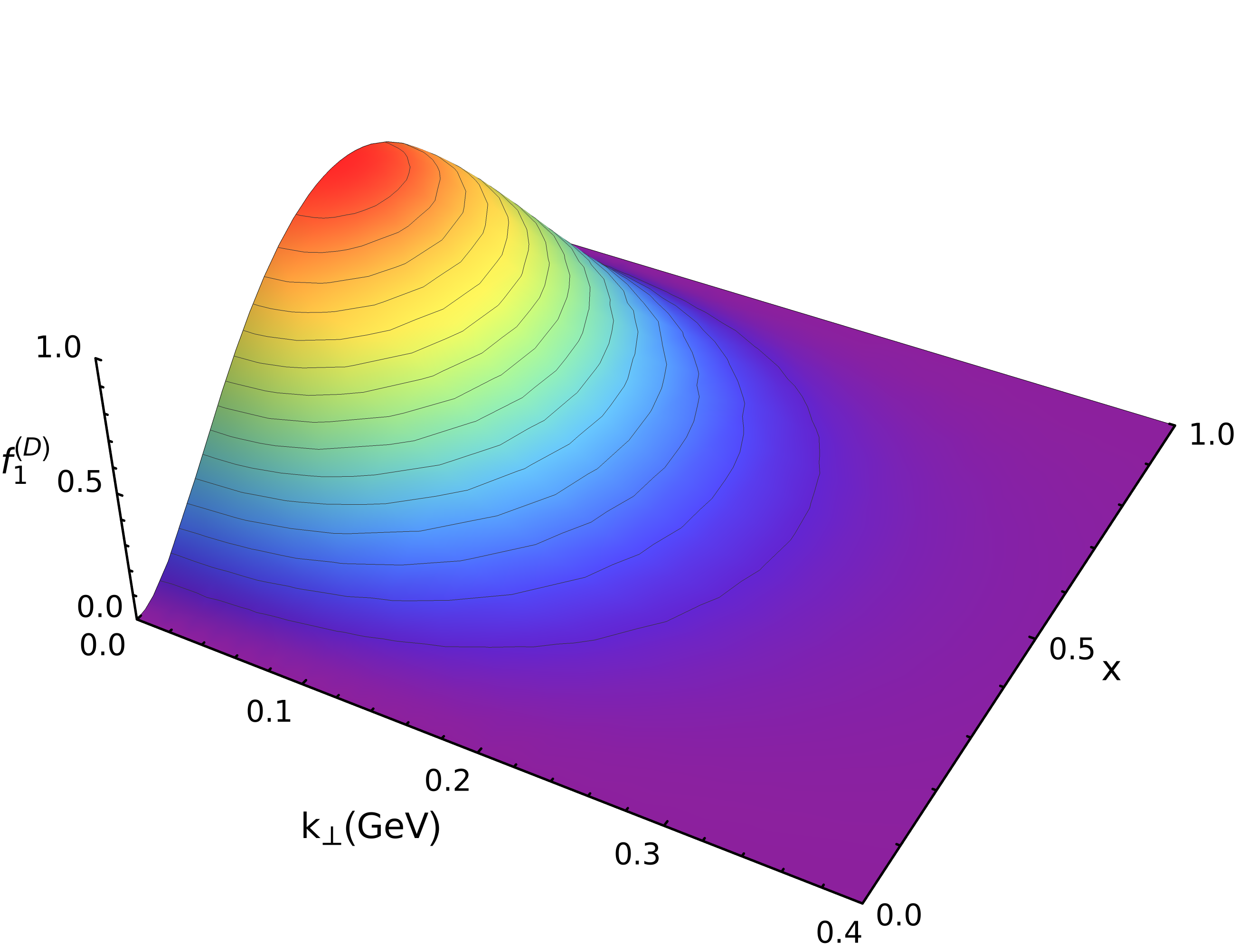}

    \caption{Unpolarized Transverse Momentum Distribution (arbitrary normalization), $f_1$, for the models (C) and (D) parametrized  with $f_\pi=130\,$MeV and  $P_{q\bar q}=70\%$. 
     Top: valence wave function (C) from Eq.~\eqref{eq:psisym} with constituent quark mass of $m_{IR}=0.344\,$GeV and mass scale $m_s=0.5\,$GeV. Bottom: running quark LF self-energy and $m_{IR}$ in the vertex function in the wave function from Eq.~\eqref{eq:psisym1} model (D) with $m_s=0.175\,$GeV. 
    }
    \label{fig:psisim3d0p2}
\end{figure}

\begin{figure}[h]
	\begin{center}
		\includegraphics[width=0.8\linewidth]{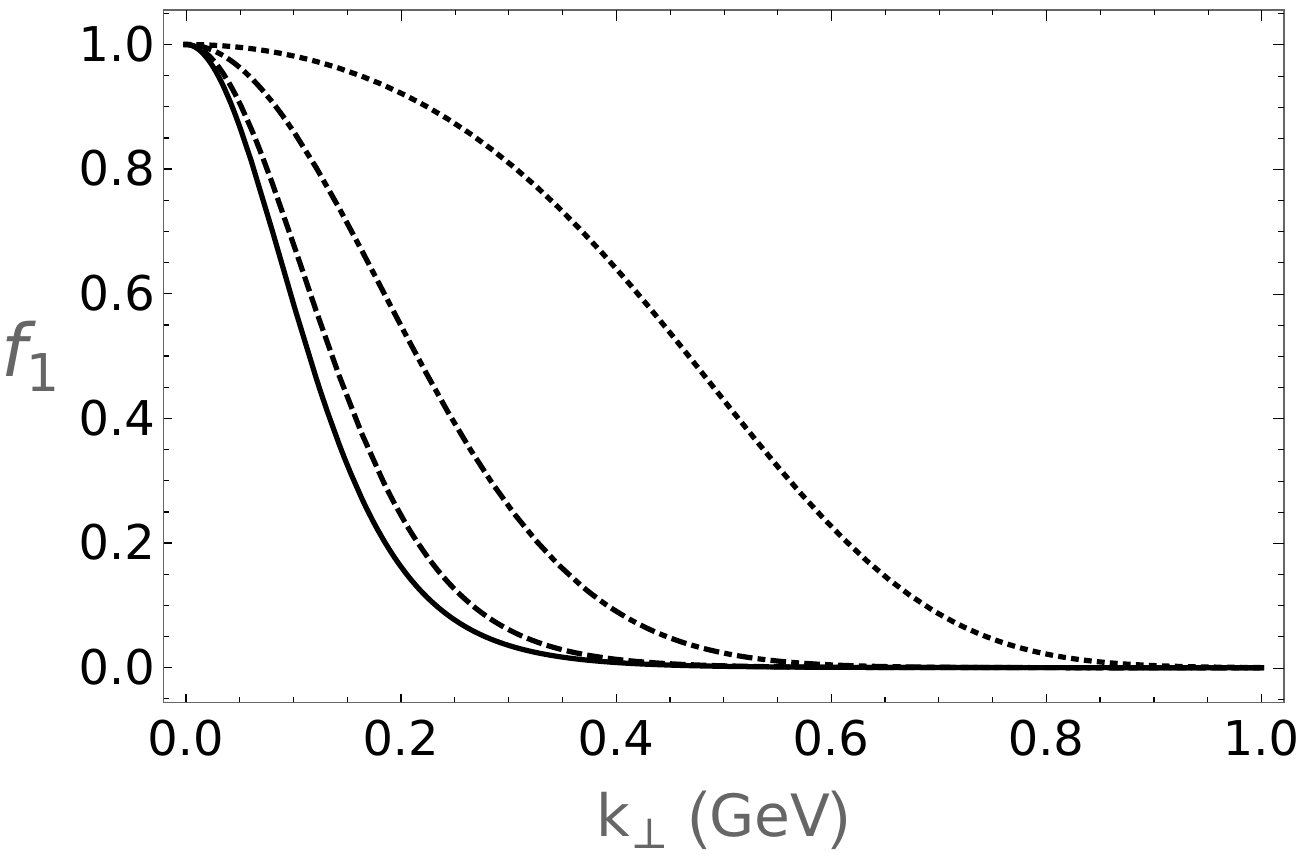}

		\caption{$f_1(k_\perp,x=0.5)$ for the valence function models. Dotted line (A),  dot-dashed line (B), dashed line (C) and solid line (D).
		}  
		\label{fig:f1kperp}
	\end{center} 
\end{figure}

\begin{figure}[t]
	\begin{center}
		\includegraphics[width=0.8\linewidth]{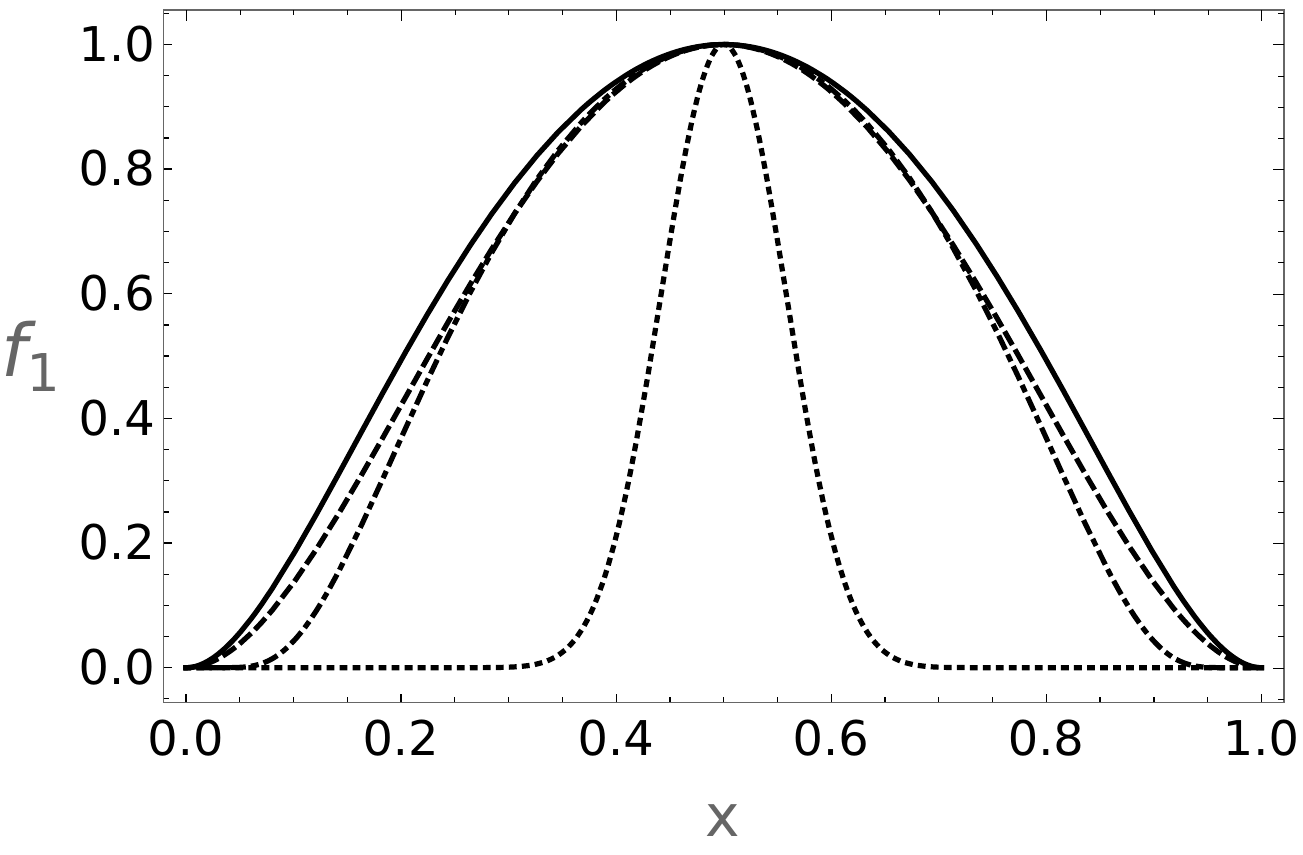}

		\caption{$f_1(k_\perp=0,x)$ for the valence function models. Dotted line (A),  dot-dashed line (B), dashed line (C) and solid line (D).
		} 
		\label{fig:f1x}
	\end{center} 
\end{figure}

\begin{figure}[h]
	\begin{center}
		\includegraphics[width=0.8\linewidth]{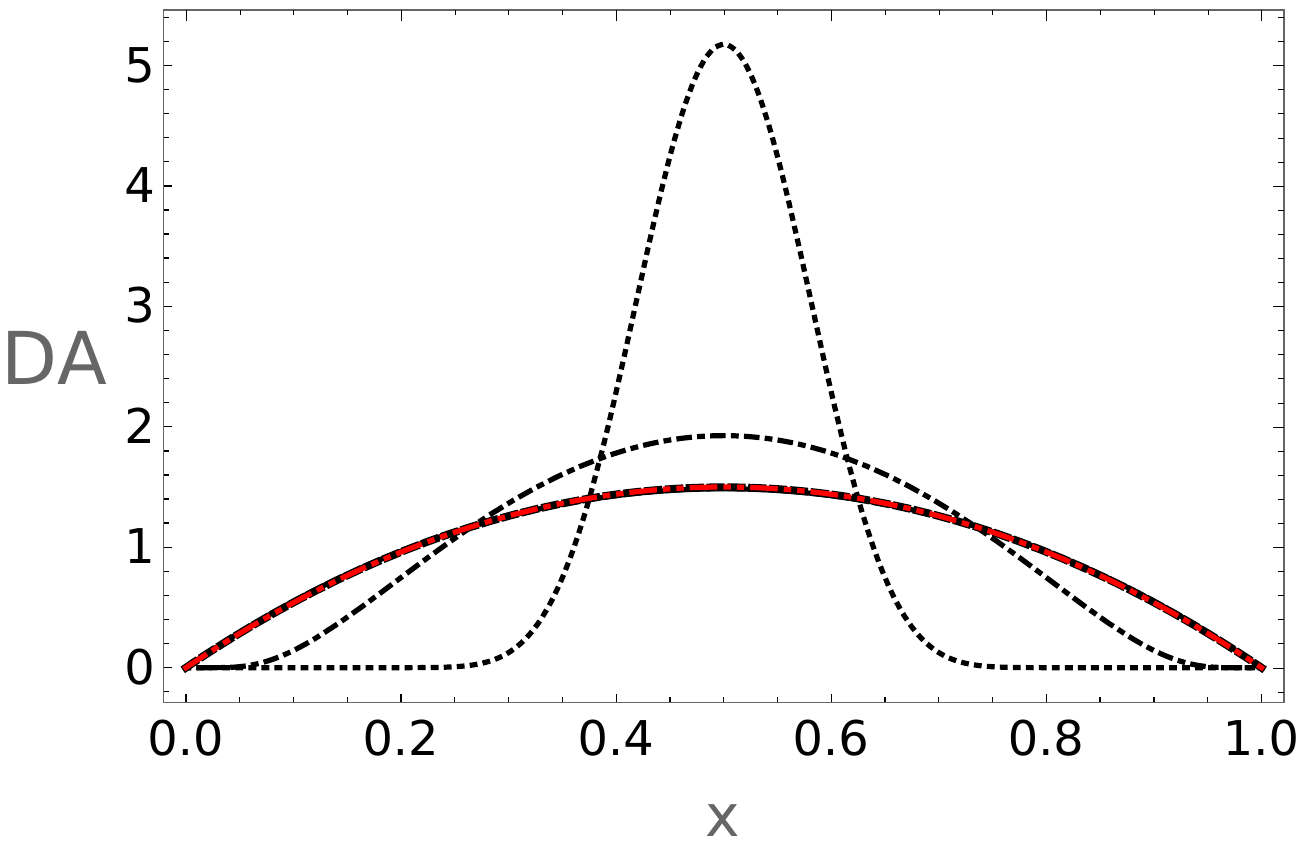}
		\caption{Pion distribution amplitude
			for the model wave function of the pion normalized to unity. Dotted line (A),  dot-dashed line (B), dashed line (C),  solid line (D) and red line asymptotic DA, namely $6x(1-x)$.
		}  
		\label{fig:DA}
	\end{center} 
\end{figure}

The unpolarized Transverse Momentum Distribution $f_1^{(c)}$ and $f_1^{(d)}$ for models (C) and (D), respectively, are shown in Fig. \ref{fig:psisim3d0p2}. These models are parametrized with $f_\pi=130\,$MeV, $P_{q\bar q}=70\%$, $m_{IR}=0.344\,$GeV resulting mass scales $m_s=0.5\,$GeV and $m_s=0.175\,$GeV for models (C) and (D), respectively. The results for $f_1\left(k_{\perp},x\right)$ for both models are very similar due to the dominance of the vertex function momentum distribution given in the square bracket in Eqs.\eqref{eq:psisym} and \eqref{eq:psisym1} over the resolvent (first term in these equations) which, in model (D), carries the running mass. Furthermore, the difference in the resolvents is also compensated by the difference in the value of $m_s$, which changes by a factor $\sim 3$ from one model to the other. Such a similarity is also present in Figs.  \ref{fig:f1kperp}, \ref{fig:f1x} and \ref{fig:DA}.

In order to detail the comparison between the models, it is shown $f_1(k_\perp,x=0.5)$ in Fig. \ref{fig:f1kperp}. The Gaussian models, (A) and (B), with and without the running mass have a quite different behavior with the transverse momentum and it is very expressive the slow decay of  
$f_1(k_\perp,x=0.5)$ for model (A) against the other models. This is also a result of the large value of $\Sigma_\text{eff}(\mathbf{k}_\perp^2,x)$ up to ${k}_\perp \sim 2\,$GeV. On the other hand, models (C) and (D) present similar results as we have already discussed. In Fig. \ref{fig:f1x} it is shown $f_1(k_\perp=0,x)$ for all valence models, where the very distinctive feature is seen for model (A) with the sharp peak around $x=0.5$, as has been observed when Fig. \ref{fig:psisim3d0p2} was discussed.

\begin{figure}[h]
	\begin{center}
		\includegraphics[width=0.8\linewidth]{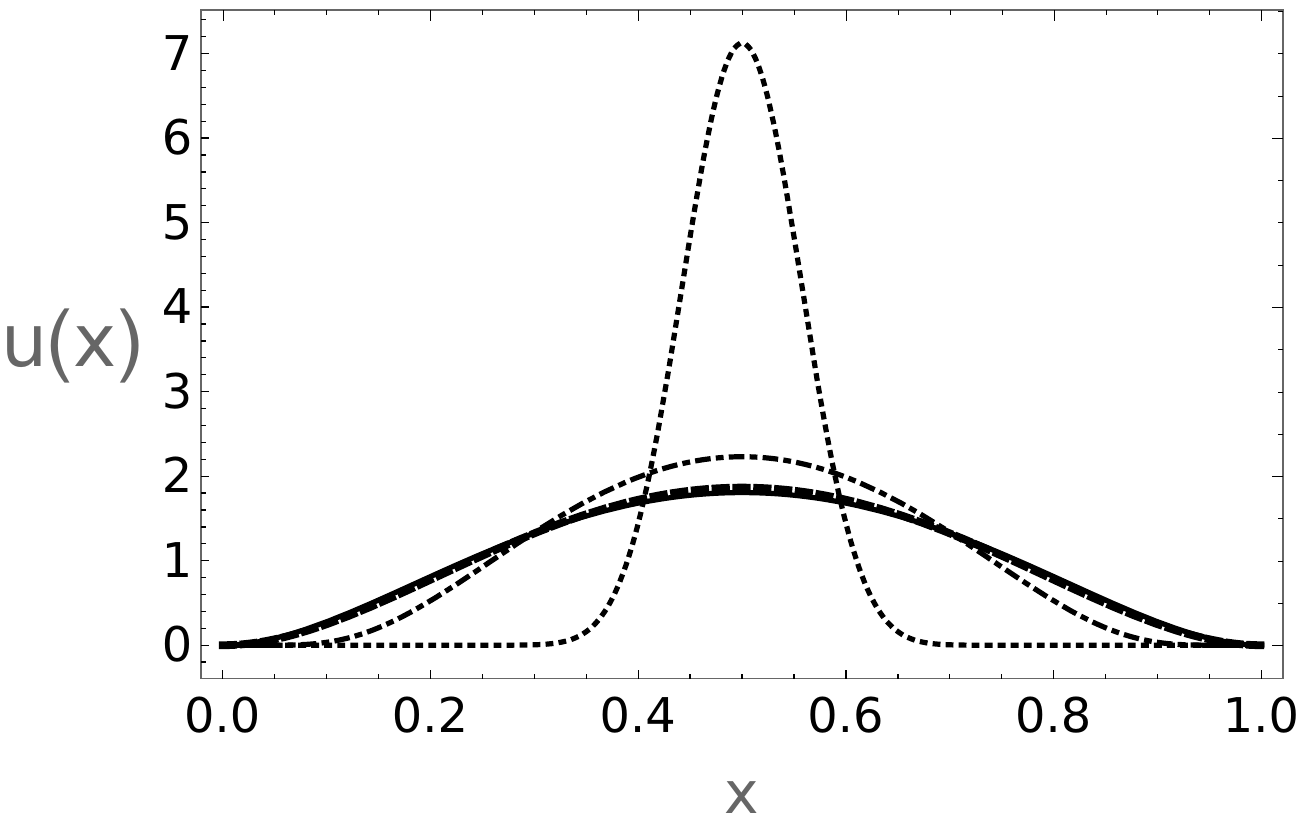}
		\caption{Pion parton distribution function, $u(x)$
			for the model wave function of the pion normalized to unity. Dotted line (A),  dot-dashed line (B), dashed line (C) and solid line (D).
		}  
		\label{fig:ux}
	\end{center} 
\end{figure}

The effect of the running quark mass in the Pion distribution amplitude is shown in Fig. \ref{fig:DA}. Again, a pronounced effect of the running mass is seen in the Gaussian model, with model $A$ sharply peaked around $x=0.5$, while models (C) and (D) have DAs essentially matching the asymptotic one, due to the polynomials structure. In Fig. \ref{fig:ux} it is shown the unpolarized Pion parton distribution function, $u(x)$
for the models (A)-(D), where, for the sake of comparison, is normalized to $1$. The features observed so far are repeated in the $u(x)$ profiles, and again the strongest effect of the running mass is seen in the Gaussian model.

\section{Summary and prospects} \label{sec:Summa}

In this work we developed an effective light-front formulation that  incorporates quark dressing effects through a momentum-dependent quark mass into the squared mass operator of light quark–antiquark systems. Starting from a Minkowski-space quark propagator constrained by a lattice-QCD–inspired parametrization, we constructed the disconnected light-front resolvent for a light quark–antiquark system by means of a generalized spectral representation, with the instantaneous contributions explicitly isolated. The resolvent was then projected onto a constituent-quark helicity basis, yielding an effective dressed mass-squared operator suitable for light-front Hamiltonian formulations and for light-front projections of the Bethe–Salpeter equation.

The resulting dressed mass operator exhibits a strong infrared enhancement, consistent with dynamical chiral symmetry breaking and nonperturbative QCD dynamics, while recovering the expected ultraviolet behavior dictated by asymptotic freedom. This behavior was encoded in an effective light-front quark self-energy, whose momentum dependence was analyzed in detail. We found that the effective quark mass reaches values well above the standard constituent quark mass scale in the infrared region and smoothly evolves toward the current quark mass at large transverse momentum.

As an application, we investigated the pion phenomenology using representative Gaussian and power-law light-front wave-function models. We computed unpolarized transverse-momentum–dependent distributions, unpolarized parton distribution functions, and distribution amplitudes, and assessed the impact of the running quark mass on these observables. The effects of quark dressing are pronounced in Gaussian models. We find that the inclusion of the effective light-front quark self-energy leads to sizable infrared modifications of the valence wave functions in the Gaussian models. For power-law–type models, the impact of the running mass is largely compensated by the vertex structure and the corresponding readjustment of the wave-function scale, resulting in similar phenomenology. Therefore, we expect major contributions of the effective light-front quark self-energies in dynamical light-front models which includes confinement.

The present formulation provides a systematic and flexible framework to embed nonperturbative quark dressing effects into light-front descriptions of hadrons. Future developments include solving the Schwinger–Dyson equation for the quark propagator in Minkowski space and  incorporating the resulting fully dressed propagator into the present light-front framework. This will allow the construction of a fully self-consistent effective dressed mass-squared operator and its application to explore the spectrum and structure of the light hadrons, such as pion, kaon and nucleon. This approach opens a promising avenue towards a unified Minkowski-space description of hadron structure, with direct relevance for hadron imaging in line with forthcoming experimental programs at future facilities such as the Electron–Ion Collider.

\acknowledgments

JPBCM and TF thanks the hospitality of the Institute of Modern Physics (IMP),  to the  President’s International Fellowship Initiative/ Chinese Academy of Sciences (PIFI/CAS) for the financial support during the visit to IMP,  to Instituto Nacional de Ci\^encia e Tecnologia - Nuclear Physics and Applications (INCT-FNA, MCTI) (Grant No. 464898/2014-5), and to Funda\c c\~ao de Amparo \`a Pesquisa do Estado de S\~ao Paulo (FAPESP) (Grant No. 2024/17816-8). JPBCM thanks the support from FAPESP (Grant
No. 2023/09539-1) and to the Conselho Nacional de Desenvolvimento Científico e Tecnológico (CNPq) (Grant No. 351403/2025-6 ). TF thanks the the support from CNPq (Grant No. 306834/2022-7) and FAPESP (Grant No. 2023/13749-1). WdP acknowledges the partial support of the CNPq (Grants No. 3313030/2021-9,
401565/2023-8, 408419/2024-5) and FAPESP (Grant 2025/05312-8). 
\begin{widetext}
\appendix

\section{Derivation of the constant $C$}\label{app:constantC}
The subtraction constant $C$ is introduced in order that the effective physical square mass operator, \eqref{eq:m02phys},  tends to the free square mass  operator with bare quark masses for large $|\mathbf{k}_\perp|$, as demanded by asymptotic freedom. In order to derive the expression for $C$ we perform the following manipulations:
\begin{eqnarray}
  C+m_0^2&=&\left\{Z_M \left[ \sum_{a,b=1}^3
  \frac{\overline \zeta_a\,\overline\zeta_b}{\mathbf{k}^2_{\perp }+ m^2_a (1-x) 
+ m^2_b\, x }\right]^{-1}\hspace{-.2cm}- \mathbf{k}_\perp^2\right\}\Bigg|_{\mathbf{k}^2_{\perp }\to\infty}\,  
\nonumber\\
&=&\mathbf{k}^2_{\perp }\left\{Z_M \left[ \sum_{a,b=1}^3
  \frac{\overline \zeta_a\,\overline \zeta_b}{1+ \frac{m^2_a (1-x) 
+ m^2_b\, x }{\mathbf{k}^2_{\perp }}}\right]^{-1}\hspace{-.2cm}- 1\right\}\Bigg|_{\mathbf{k}^2_{\perp }\to\infty}\,
\nonumber\\
&=&\mathbf{k}^2_{\perp }\left\{Z_M \left[ \sum_{a,b=1}^3
  {\overline \zeta_a\,\overline\zeta_b}{\Big(1- \frac{m^2_a (1-x) 
+ m^2_b\, x }{\mathbf{k}^2_{\perp }}\Big)}\right]^{-1}\hspace{-.2cm}- 1\right\}\Bigg|_{\mathbf{k}^2_{\perp }\to\infty}\,
\nonumber \\
&=&\mathbf{k}^2_{\perp }\left\{ \left[ 1 - \sum_{a,b=1}^3
  {\overline\zeta_a\,\overline\zeta_b}{\Big(\frac{m^2_a (1-x) 
+ m^2_b\, x }{Z_M\,\mathbf{k}^2_{\perp }}\Big)}\right]^{-1}\hspace{-.2cm}- 1\right\}\Bigg|_{\mathbf{k}^2_{\perp }\to\infty}\,
\nonumber \\
&=&  \sum_{a,b=1}^3
{\overline\zeta_a\,\overline\zeta_b}{\frac{m^2_a (1-x) 
+ m^2_b\, x }{Z_M}}\, \nonumber\\
&=& \sum_{a,b=1}^3
{\overline\zeta_a\,\overline\zeta_b}{\frac{m^2_a 
 }{Z_M}} \,,
\end{eqnarray}
and from Eq.~\eqref{eq:Z} that gives $Z_M^\frac12=\sum_{b=1}^3\overline \zeta_b $ we find that:
\begin{equation}\label{eq:Cfinal_app}
C= Z_M^{-\frac12} \sum_{a=1}^3  \overline \zeta_a\,m^2_a -m^2_0\,.
\end{equation}
which is Eq.~\eqref{eq:Cfinal} in Sect.~\ref{sec:ELCQSE}.

\end{widetext}

\bibliography{ref}

\end{document}